\documentclass[onecolumn,authoryear]{els-mrw} 

\usepackage{amsmath,amssymb,amsfonts,amsthm,makeidx,graphicx}
\usepackage{txfonts}
\usepackage{helvet}
\usepackage{bm}
\usepackage{color}
\usepackage{braket}
\usepackage{physics}

\makeatletter
\providecommand{\tocchapnumskip}{2.5em}
\renewcommand{\tocchapnumskip}{2.5em}
\makeatother

\makeatletter
\newcommand{\l@chapter}{\@dottedtocline{1}{0pt}{2.5em}} 
\providecommand{\tocchapnumskip}{2.5em}
\renewcommand{\tocchapnumskip}{2.5em}
\makeatother

\usepackage{tocloft}

\renewcommand{\braket}[1]{%
  \left\langle #1\right\rangle
}

\def\bnab{\mathrel{{ \mathbf{\nabla}}}}
\def\bsig{\mathrel{{ \mathbf{\sigma}}}}

\begin{document}

\chapter{Higher-order corrections to beta decays and nuclear structure}\label{chap1}

\author[1]{Grigor Sargsyan}%
\author[2]{Ayala Glick-Magid}%

\address[1]{\orgname{Michigan State University, \orgdiv{Facility for Rare Isotope Beams}}, \orgaddress{640 S Shaw Ln, East Lansing, MI 48864, USA}}
\address[2]{\orgname{Tel-Aviv University}, \orgdiv{School of Physics and Astronomy}, \orgaddress{ Tel-Aviv 69978, Israel}}

\articletag{Chapter Article tagline: update of previous edition,, reprint..}

\maketitle




\begin{glossary}[Nomenclature]
\begin{tabular}{@{}lp{34pc}@{}}
BSM & beyond the Standard Model\\
CKM & Cabibbo–Kobayashi–Maskawa\\
CVC & conserved vector current \\
LO & Leading order \\
NLO & Next to leading order \\
PDG & Particle Data Group \\
$V-A$ & vector minus axial-vector
\end{tabular}
\end{glossary}

\begin{abstract}[Abstract]
Precision measurements of nuclear $\beta$ decay offer a powerful route to tests of physics beyond the Standard Model, but interpreting measurements spectra and correlation observables requires control of subtle nuclear-structure effects. This chapter develops a framework for calculating higher-order corrections--small contributions that can become important at the precision sought in searches for new interactions. After motivating their role in $\beta$-decay observables (with radiative corrections treated in the accompanying chapter), we introduce the relevant kinematics and multipole expansion of the shape factor, identify the subleading contributions, and present the impulse-approximation multipole operators. We then describe the nuclear many-body methods used to obtain the required states and transition densities, and derive the corresponding operator matrix elements in the harmonic-oscillator basis. Taken together, these ingredients enable systematic calculations of nuclear shape factors beyond the long-wavelength approximation, including higher-order multipoles where needed. By helping distinguish Standard-Model recoil effects from possible signatures of new physics, this framework supports the interpretation of increasingly precise 
$\beta$-decay experiments.
\end{abstract}

\tableofcontents
\markboth{}{} 

\section{Introduction}

Precision studies of nuclear $\beta$ decay provide a sensitive low-energy window onto the structure of the weak interaction and possible physics beyond the Standard Model (BSM). Although searches for new physics are often associated with high-energy facilities such as the Large Hadron Collider, low-energy measurements of radioactive nuclei offer a complementary approach. Their sensitivity arises from the combination of precisely measured decay rates, energy spectra, and angular correlations with the comparatively clean theoretical structure of the weak interaction. In several cases, nuclear $\beta$-decay experiments provide constraints on new interactions that are competitive with those obtained at high-energy colliders.

Nuclear $\beta$ decay also plays a central role in determining parameters of the Standard Model. At the quark level, the process is mediated by charged
$W$-boson exchange between a lepton current and a light-quark current. The coupling of the latter to the charged weak interaction is described by the Cabibbo--Kobayashi--Maskawa (CKM) matrix
[\cite{Cabibbo:1963yz,Kobayashi:1973fv}]. In particular, superallowed nuclear $\beta$ decays provide the most precise determination of the matrix element $V_{ud}$ [\cite{PDG2020}]. Together with other determinations of CKM-matrix elements, this information enables stringent tests of CKM unitarity. These tests have received renewed attention following recent indications of a possible deviation from unitarity at approximately the $3\sigma$ level [\cite{cirigliano2023scrutinizing}], including scenarios involving right-handed quark couplings
[\cite{Cirigliano:2023nol}].

The same measurements test the chiral structure of the charged-current interaction. In the Standard Model, the relevant quark and lepton currents have the left-handed vector-minus-axial-vector structure, commonly denoted by $V-A$. More general effective interactions can contain right-handed vector and axial-vector, scalar, pseudoscalar, and tensor components. Such terms may be generated by heavy new particles that are not directly accessible at the energies of the decay, but their effects can be observed through small deviations in measurable decay observables. Examples include the electron--antineutrino angular correlation, $\beta$-particle asymmetries, other angular and momentum correlations, and the detailed shape of the $\beta$ spectrum. Precision measurements of these quantities therefore provide complementary tests of the Standard Model and sensitive probes of exotic scalar and tensor couplings
[\cite{Severijns2006,VosWT2015,gonzalez2019new,Falkowski2021}].

The interpretation of such measurements requires a theoretical description that is commensurate with the experimental precision. At leading order, allowed $\beta$ decay is described in terms of the Fermi and Gamow--Teller operators. This approximation captures the dominant transition amplitudes, but it is not sufficient for sub-percent tests of the weak interaction. Several classes of higher-order effects must be included, including electromagnetic radiative corrections, isospin-breaking contributions, electrostatic effects, and recoil-order corrections. The need to control these effects is common to precision programs in nuclear physics more broadly
[\cite{Cirigliano:2013xha,naviliat2013prospects,VosWT2015,Falkowski2021, BurkeySGS2022, Brodeur2023nuclear1}]. An incomplete or imprecise treatment can shift the extracted Standard-Model parameters or generate distortions that imitate the effects of BSM interactions.

Radiative corrections arise from electromagnetic interactions involving the charged particles and the weak transition. They include, among other effects, virtual-photon contributions, real-photon emission, and short-distance electroweak corrections. These contributions are essential for precision determinations of quantities such as $V_{ud}$ and for the consistent interpretation of correlation measurements. A detailed discussion of radiative corrections is given in the corresponding chapter of this volume; they will therefore be mentioned here only where they overlap with the recoil-order analysis.


The primary focus of this chapter is on recoil-order corrections. They originate from the finite mass of the nucleons and from subleading terms in the nuclear weak current. In a low-momentum expansion, their characteristic scale is set by powers of $q/m_N$, where $q$ is the momentum transferred to the nucleus and $m_N$ is the nucleon mass. Since typical $\beta$-decay momentum transfers are of order a few MeV$/c$, recoil effects are usually suppressed relative to the leading Fermi and Gamow--Teller contributions. Nevertheless, for measurements with uncertainties at or below the percent level, this suppression is not sufficient to justify neglecting them. Recoil terms can also become relatively more important when the leading amplitude is small, when different contributions interfere, or when a particular observable is especially sensitive to subleading currents. 

The recoil-order contributions contain both kinematic and dynamical
information. Kinematic effects result from expanding the nuclear
currents in inverse powers of the nucleon mass. Dynamical contributions are
associated with subleading form factors and multipole operators, including
weak magnetism, induced tensor terms, induced pseudoscalar contributions, and
higher-order axial-vector operators. Through their interference with the
leading Fermi or Gamow--Teller amplitudes, these terms modify the energy
dependence of the $\beta$ spectrum, the angular-correlation coefficients, and
the asymmetries measured in polarized decays. Although their typical size is
at the $10^{-3}$--$10^{-2}$ level, their impact can be comparable to the
effects sought in precision searches for nonstandard scalar, tensor, or
right-handed interactions.

A reliable treatment of recoil-order effects is consequently necessary for
two related reasons. First, these corrections are part of the Standard-Model
prediction against which experimental data must be compared. Second, their
uncertainties determine how confidently an apparent deviation can be
interpreted as evidence for BSM physics rather than as an incompletely known
nuclear-structure effect. This issue is particularly important for
correlation measurements and spectral-shape analyses, where recoil
contributions can have energy or angular dependencies similar to those
generated by exotic currents.

Recent progress in nuclear \emph{ab initio} theory has opened the possibility
of calculating recoil-order corrections with controlled uncertainties for an
increasing range of nuclei. Such methods aim to describe the nuclear states
and weak transition operators starting from realistic nuclear interactions, while systematically improving the many-body treatment as the interaction
and computational truncations are refined. Recent applications include
calculations based on several complementary many-body approaches
[\cite{glick2022nuclear,Sargsyan_A8,King:2022zkz,LongfellowGSB2024,
Gennari:2024sbn,Cirigliano:2024rfk,Cirigliano:2024msg,King:2025fph, sargsyan2026role, glick2026ab1}].
The systematic-improvement properties of \emph{ab initio} calculations are discussed in
[\cite{Hergert:2020bxy,Ekstrom:2022yea}]. These developments are particularly valuable because they allow nuclear recoil effects to be connected directly to the underlying nuclear dynamics rather than treated
solely through phenomenological estimates.

This chapter introduces the role of recoil-order corrections in precision nuclear $\beta$ decay and places them in the broader context of tests of fundamental symmetries. We first summarize the structure of the weak transition and the observables used in precision measurements. We then introduce the multipole representation of the shape factor and its extension to nonstandard interactions.
Using an expansion in the relevant small parameters,
we identify the subleading operators and form factors, and discuss recoil-order and nuclear-structure-dependent electrostatic contributions to decay spectra and correlation coefficients. Finally, we review modern nuclear-structure calculations, with emphasis on \emph{ab initio} approaches, and consider the theoretical developments required to match the precision of current and
future experiments.

\section{Recoil-order effects in nuclear beta decay}
Historically, precision studies of angular observables in nuclear $\beta$ decay
played a central role in establishing the $V-A$ structure of the charged weak current ~[\cite{lee1956question,jackson1957possible}]. These observables remain important tools in contemporary searches for physics beyond the Standard Model (BSM). In particular, extensions of the Standard Model may generate additional Lorentz structures, including scalar (S), pseudoscalar (P), and tensor (T) couplings. Measurements of the electron--neutrino angular correlation, the $\beta$-particle asymmetry, and the Fierz interference coefficient therefore provide complementary constraints on such interactions. A new generation of experiments is expected to improve the precision of these observables and thereby increase their sensitivity to small nonstandard contributions
[\cite{gonzalez2019new,Brodeur2023nuclear1}].

In standard treatments of $\beta$ decay, recoil-order contributions are often neglected because they are suppressed by powers of the ratio $q/m_N$, where $q$ is the momentum transferred to the nucleus and $m_N$ is the nucleon mass~[\cite{Holstein:1974zf}]. Since typical momentum transfers are of
order a few MeV$/c$, these effects are usually below the percent level
relative to the leading Fermi and Gamow--Teller (GT) amplitudes. Their
suppression, however, does not make them irrelevant in high-precision
measurements. Once experimental uncertainties become comparable to the
recoil scale, these terms must be included in the theoretical description.
This is especially important for transitions in which the leading amplitude
is reduced or for observables in which recoil contributions are enhanced by
kinematics or interference effects.

Recoil-order corrections consequently provide a valuable probe of
sub-percent physics in allowed $\beta$ decay. If they are not properly
controlled, they can produce distortions that resemble signals of BSM
interactions. The relevant contributions originate from subleading pieces of
the nuclear weak current, including weak magnetism, induced tensor terms, and
higher-rank axial-vector multipoles. Through interference with the leading
operators, these contributions modify both the $\beta$-energy spectrum and
the associated angular correlations, typically at the
$\mathcal{O}(10^{-3})$ level. A reliable calculation of these corrections
within \emph{ab initio} many-body methods is therefore necessary for
separating genuine BSM effects from Standard-Model contributions generated by
nuclear structure.

\subsection{$\beta$-decay kinematics and the basic differential rate}
\label{sec:kinematics}

Nuclear $\beta$ decay is the process in which a nucleus emits a charged lepton, an electron or a positron, together with the corresponding electron-flavor
(anti)neutrino. The mass number of the nucleus, $A$, remains unchanged, while the atomic number $Z$ changes by one unit. 
In $\beta^-$-decay,
\begin{equation*}
^A_Z X \longrightarrow   {}_{Z+1}^{\quad A} Y + e^- + \bar{\nu}_e,
\end{equation*}
the atomic number increases by one, whereas in $\beta^+$-decay,
\begin{equation*}
^A_Z X \longrightarrow   {}_{Z-1}^{\quad A} Y + e^+ + \nu_e,
\end{equation*}
it decreases by one.
At the fundamental level, $\beta$ decay is mediated by the weak interaction, which is much weaker than the strong interaction between nucleons and the electromagnetic interaction at nuclear scales.
The weakness of this interaction allows the transition amplitude to be evaluated perturbatively, to leading order in the weak Hamiltonian. The transition rate for $\beta$-decay into a continuum of final states is then obtained from Fermi's golden rule\footnote{Throughout this section, we use $\hbar=c=1$ units.}:
\begin{equation}
d\Gamma = 2\pi |M_{fi}|^2 \delta(E_i - E_f) \, d\rho_f,
\label{eq:golden}
\end{equation}
where $\delta$ is the Dirac delta function, $E_i$ and $E_f$ are the energies of the initial and final states, respectively, $d\rho_f$ is a density element of the final state, and 
\begin{equation}
   M_{fi}= \braket{\phi_f|H'|\phi_i}
\end{equation}
is the transition matrix element, with $H'$ being the perturbing weak Hamiltonian, and $\phi_i$ and $\phi_f$, the initial and final state wavefunctions.

The initial state simply involves the initial nucleus and can be considered stationary.
The final state contains three particles: a charged lepton, a neutral lepton, and the final nucleus. Since final-state interactions of the neutrino are negligible, its wavefunction can be approximated with plane waves, $\psi_\nu(r) = (1/\sqrt[]{V})\exp(i\vec{k_\nu} \cdot \vec{r})$, where $1/\sqrt[]{V}$ is the normalization constant and $\vec{k}_\nu$ is the momentum of the neutrino. Moreover, if we temporarily neglect the interaction between the electron (positron) and the Coulomb field of the nucleus, which will be restored below, then the wavefunction of the charged lepton can also be expressed by plane waves $\psi_e(r) = (1/\sqrt[]{V})\exp(i\vec{k}_e \cdot \vec{r})$, where $\vec{k}_e$ is the momentum of the charged lepton. As typical $\beta$-decay energies are of a few MeV and nuclear radii $R$ are of a few fm, in standard derivations (see, for example, \cite{wong2008introductory}), the product of the two plane waves, $\psi(r) = (1/{V})\exp(i\vec{q} \cdot \vec{r})$, with $\vec{q}=\vec{k}_\nu+\vec{k}_e$, is often expanded in $qr \ll 1$, retaining only the leading term which corresponds to allowed $\beta$ decays. 

In the Standard Model, the semileptonic weak Hamiltonian can be written as (see, \textit{e.g.}, \cite{walecka2004theoretical})
\begin{equation}
    H'=-\frac{G_FV_{ud}}{\sqrt{2}}\int d^3r j_\mathrm{lept}^{\,\mu}(\vec{r})\mathcal{J}_\mu(\vec{r}),
\end{equation}
where $G_F$ is the Fermi weak coupling constant, $V_{ud}$ is the CKM matrix element, $j^{\,\mu}_{\mathrm{lept}}(\vec{r})$ and $\mathcal{J}_\mu(\vec{r})$ are leptonic and hadronic (nuclear) currents, respectively (the forms of these currents in terms of Dirac spinors can be found in, \textit{e.g.}, \cite{gonzalez2019new} or \cite{Hayen2026}), and we use the metric $g_{\mu\nu}=\operatorname{diag}(1,-1,-1,-1)$. Hence, the transition matrix element is given by
\begin{equation}
    M_{fi}=-\frac{G_FV_{ud}}{\sqrt{2}V}l^{\,\mu}\int d^3r \, e^{i\vec{q}\cdot\vec{r}} \bra{f} \mathcal{J}_\mu(\vec{r}) \ket{i},
\end{equation}
where we used $j_\mathrm{lept}^{\,\mu}=l^{\,\mu}{e^{i\vec{q}\cdot \vec{r}}}/{V}$, assuming plane-wave lepton states. When evaluating the transition probability for an unpolarized initial state, we must average over the initial nuclear spin states $(2J_i + 1)$ and sum over final magnetic substates and lepton spins. With this and by denoting the coordinate-dependent part by the shape factor $\Theta(\vec{q}, \vec{\beta}\cdot\hat{\nu})$, the square of the transition matrix element $|M_{fi}|^2$ can be expressed as:
\begin{equation}
\label{eq:transition_matrix_elemen}
|M_{fi}|^2 = \frac{G_F^2 V_{ud}^2}{V^2}   \frac{4\pi}{2J_i + 1} \Theta(\vec{q}, \vec{\beta}\cdot\hat{\nu}),
\end{equation}
where $\vec{\beta}=\vec{k}_e/E_e$ and $\hat{\nu}=\vec{k}_{\nu}/|\vec{k}_{\nu}|$ denote the charged-lepton velocity in natural units and the unit vector in the neutrino direction, respectively. The additional factor of $4\pi$ is separated from $\Theta$ for convenience, while 2 in the denominator is absorbed into it. We will see that the $4\pi$ factor will cancel when one derives the leading allowed beta decay rate using the long-wavelength approximation. 

To obtain the decay rate in Eq. \eqref{eq:golden}, we also need the density element of the final state. The final state contains three particles. However, because the final nucleus is much heavier than the leptons, it carries the recoil momentum required by momentum conservation without taking away significant kinetic energy (possible corrections from nuclear recoil will be discussed below). The total available kinetic energy is shared between the electron and the neutrino, $E_0 = E_e + E_\nu$, where $E_0$ is the endpoint energy.

Assuming the decay occurs in a normalization volume $V$, the number of states for a free neutrino of momentum $k_\nu$ is given in statistical mechanics to be
\begin{equation}
    d\rho_\nu=\frac{V d^3{k}_\nu}{(2\pi)^3}. 
\end{equation}
We now restore the interaction between the charged lepton and the Coulomb field of the final nucleus, which was temporarily neglected above. As a result, the electron (positron) is no longer a free particle, and its phase-space contribution is multiplied by the Fermi function,  which accounts for the leading point-Coulomb distortion of the charged-lepton wavefunction:
\begin{equation}
    d\rho_e=\frac{V d^3{k}_e}{(2\pi)^3} F_0^{\mp}(Z_f,E_e),
\end{equation}
where $Z_f$ is the charge of the final nucleus. The Fermi function for $\beta^\mp$ decay is given by~\cite{fermi1934} 
\begin{equation}
F_0^{\mp}(Z_{f},E_e) = \frac{4\,(2k_e R_{f})^{2(\gamma_{0}-1)}\,\bigl|\Gamma(\gamma_{0}+iy^\mp)\bigr|^{2}}{\bigl[\Gamma(2\gamma_{0}+1)\bigr]^{2}} \;e^{\pi y^\mp}
\end{equation}
with
\begin{equation*}
\gamma_{0} = \sqrt{1-(\alpha Z_{f})^{2}}, \qquad y^\mp = \pm \alpha Z_{f} \frac{E_e}{k_e},
\end{equation*}
where the upper sign in $y$ is for $\beta^{-}$ decay and the lower sign is for $\beta^{+}$ decay. Here, $R_{f}$ is the charge radius of the final nucleus, $\alpha$ is the fine-structure constant, and $\Gamma(x)$ is the Euler gamma function.

Therefore, the total density element of the final state is given by:
\begin{equation}
    d\rho_f = \frac{V d^3 k_e}{(2\pi)^3} \frac{V d^3 k_\nu}{(2\pi)^3} F_0^\mp(Z_f,E_e)\delta(E_0 - E_e - E_\nu). 
\end{equation}
We express the momentum volume elements in spherical coordinates ($d^3{k} = k^2 dk d\Omega$), and use the relativistic energy-momentum relation $E^2 = k^2 + m^2$ to obtain $k dk = E dE$. For the electron, this gives
$d^3{k}_e = k_e E_e dE_e d\Omega_e$, whereas for the neutrino (treating it as effectively massless, so that $E_\nu = k_\nu$) we have $d^3{k}_\nu = E_\nu^2 dE_\nu d\Omega_\nu$. Substituting these expressions back into the phase-space density yields
\begin{equation}
d\rho_f = \frac{V^2}{(2\pi)^6} k_e E_e dE_e d\Omega_e E_\nu^2 dE_\nu d\Omega_\nu F_0^\mp(Z_f,E_e) \delta(E_0 - E_e - E_\nu).
\end{equation}
Integrating over the neutrino energy $E_\nu$ using the Dirac delta function fixes the conservation condition $E_\nu = E_0 - E_e$, yielding
\begin{equation}
\frac{d\rho_f}{dE_e d\Omega_e d\Omega_\nu} = \frac{V^2}{(2\pi)^6} k_e E_e (E_0 - E_e)^2 F_0^\mp(Z_f,E_e).
\end{equation}

Finally, combining the density of states obtained with the transition matrix element from Eq. \eqref{eq:transition_matrix_elemen} into Eq. \eqref{eq:golden} yields the full differential decay rate:
\begin{equation}
\frac{{d}^5\Gamma}{{d}E_e {d}\Omega_e {d}\Omega_\nu} = \frac{G_F^2 V_{ud}^2}{(2\pi)^5} (E_0 - E_e)^2 k_e E_e F_0^\mp(Z_f, E_e) C_{\text{corr}} \frac{4\pi}{2J_i + 1} \Theta(\vec{q}, \vec{\beta}\cdot\hat{\nu}).
\label{eq:dec_rate}
\end{equation}
The correction factor $C_{\text{corr}}$ incorporates radiative and outer kinematic corrections independent of the nuclear core structure, such as atomic exchange, atomic mismatch (including shake-up and shake-off), atomic screening and molecular screening (see, \textit{e.g.}, \cite{Hayen:2018}). One such is the recoil of the nucleus.
%
If it is included in the derivation of the decay rate, then the expression in Eq. \eqref{eq:dec_rate} needs to be multiplied by the recoil correction factor:
\begin{equation}
    W_\mathrm{rec}= \left( 1- \frac{\Delta M}{M_i}\right) \left[ 1+ \frac{E_e}{M_i}(1-\vec\beta\cdot \hat\nu)+ \mathcal{O}(\frac{(\Delta M)^2}{M_{\min}^2})\right].
    \label{eq:recoil_nucleus}
\end{equation}
For a nucleus of $A=10$ those corrections are of the order of $\sim10^{-3}$. 

Another correction which will be described here is the finite nuclear charge distribution, which produces electrostatic
corrections beyond the point-charge Fermi function.
At leading order, a contribution proportional to the
uncorrected shape factor can be included through the
replacement~[\cite{benatar2026leading}]
\begin{equation}
F_0^\mp(Z_f,E_e)
\longrightarrow
F_0^\mp(Z_f,E_e)+\Delta F^\mp_{\mathrm{EC}},
\qquad
\Delta F^\mp_{\mathrm{EC}}
=\mp\frac{4}{\sqrt{6}}
\alpha Z_f E_e R_f,
\label{eq:electrostatic_Fermi}
\end{equation}
where the charge radius of the final nucleus, $R_{f}$, should be taken here as the rms charge radius. As before, the upper (lower) sign corresponds to
$\beta^-$ ($\beta^+$) decay.
This contribution is present for arbitrary Standard Model
$\beta$ transitions.

An additional electrostatic correction is due to the difference between the Coulomb potentials of the decaying and final nucleus, $\Delta E_{\mathrm c}$, that, for a $\beta^-$ ($\beta^+$) decay, should be added to (subtracted from) the maximal electron (positron) energy $E_0$ arising from temporal derivations of the nuclear current introduced into $\Theta$~[\cite{behrens1982electron}] (we will present this correction explicitly in the expressions in Subsec. \ref{subsec:multipole_impolse}). The approximation of this Coulomb difference for a homogeneously charged sphere is $\frac{6}{5}\frac{\alpha Z_{\text{f}}}{R}$, but for precision calculations, it is usually better using the experimental value of the Coulomb displacement energy between a pair of isobaric analog levels, which is given by $\Delta E_{\mathrm c} \equiv M_a\left(Z_>\right)-M_a\left(Z_<\right)+\Delta_{nH}$, where $M_a\left(Z_>\right)$ ($M_a\left(Z_<\right)$) is the atomic mass of the larger (smaller) Z member of the analog pair, and $\Delta_{nH}$ is the neutron-hydrogen mass difference~[\cite{ANTONY19971}]. These can be found in the literature with high accuracy for most nuclei and states (see, \textit{e.g.},~\cite{ANTONY19971,osti_1774904}).

As noted above, the nuclear-structure contributions associated with the weak current are contained in the shape factor $\Theta\left(q,\vec{\beta}\cdot\hat{\nu}\right)$,
thus, reducing the whole problem of recoil-order effects to a systematic expansion of $\Theta$.

\subsection{Multipole expansion of the shape factor}

A convenient and systematic treatment of nuclear-structure effects is obtained by
expanding the weak nuclear current in multipole operators. Several equivalent
conventions have been developed in the literature. Below we present the formalism by
Donnelly--Walecka [\cite{walecka2004theoretical}]. This approach was adapted for high-precision
applications in~\cite{glick2022formalism}, and it applies to arbitrary
$\beta$ transitions and can be extended to any desired order in the relevant
kinematic expansion.

In this notation, the shape factor for a generic $\beta^{\mp}$ decay is given
by [\cite{walecka2004theoretical}]
\begin{multline}
\Theta\left(q,\vec{\beta}\cdot\hat{\nu}\right)=
\sum_{J=1}^{\infty}\left[\left(1-\left(\hat{\nu}\cdot\hat{q}\right)\left(\vec{\beta}\cdot\hat{q}\right)\right)\left(|\braket{\Vert \hat{E}_{J}\Vert} |^{2}+|\braket{\Vert \hat{M}_{J}\Vert} |^{2}\right)
\pm\hat{q}\cdot\left(\hat{\nu}-\vec{\beta}\right)
2\mathfrak{Re}\left(\braket{\Vert \hat{E}_{J}\Vert} \braket{\Vert \hat{M}_{J}\Vert}^{*}\right)\vphantom{\left|\left\langle \left\Vert \hat{M}_{J}\right\Vert \right\rangle \right|^{2}}\right]\\
+\sum_{J=0}^{\infty}\left[\left(1-\vec{\beta}\cdot\hat{\nu}+2\left(\hat{\nu}\cdot\hat{q}\right)\left(\vec{\beta}\cdot\hat{q}\right)\right)|\braket{\Vert \hat{L}_{J}\Vert} |^{2}+\left(1+\vec{\beta}\cdot\hat{\nu}\right)|\braket{\Vert \hat{C}_{J}\Vert} |^{2}
-\hat{q}\cdot\left(\hat{\nu}+\vec{\beta}\right)
2\mathfrak{Re}\left(\braket{\Vert \hat{L}_{J}\Vert} \braket{\Vert \hat{C}_{J}\Vert}^{*}\right)\vphantom{\left|\left\langle \left\Vert \hat{M}_{J}\right\Vert \right\rangle \right|^{2}}\right]\text{,}\\
\label{eq:multipole_exp}
\end{multline}
where the upper (lower) sign corresponds to
$\beta^-$ ($\beta^+$) decay, and $\braket{\Vert \hat{O}_{J}\Vert}$ denotes the reduced matrix element\footnote{We use the following convention for the Wigner-Eckhart theorem $\langle J_f M_f | \hat O_{JM} | J_i M_i \rangle = (-1)^{J_f - M_f} \begin{pmatrix} J_f & J & J_i \\ -M_f & M & M_i \end{pmatrix} \langle J_f \parallel \hat O_J \parallel J_i \rangle
$.} of
a rank-$J$ spherical tensor $\hat{O}_{J}$ between the final and initial nuclear states. The tensor is constructed from the
axial-vector or polar-vector part of the nuclear current and evaluated
between the initial and final nuclear states.

The four multipole operators entering Eq.~\eqref{eq:multipole_exp} are the
Coulomb, longitudinal, electric, and magnetic multipoles:
\begin{align}
\hat{C}_{JM}\left(q\right)
&\equiv
\int d^{3}r\,
S_{JM}\left(q\vec{r}\right)\mathcal{J}_{0}\left(\vec{r}\right),\\
\hat{L}_{JM}\left(q\right)
&\equiv
\frac{i}{q}\int d^{3}r\,
\left[\vec{\nabla}S_{JM}\left(q\vec{r}\right)\right]
\cdot\vec{\mathcal{J}}\left(\vec{r}\right),\\
\hat{E}_{JM}\left(q\right)
&\equiv
\frac{1}{q}\int d^{3}r\,
\left[\vec{\nabla}\times
\vec{S}_{JJ1}^M\left(q\vec{r}\right)\right]
\cdot\vec{\mathcal{J}}\left(\vec{r}\right),\\
\hat{M}_{JM}\left(q\right)
&\equiv
\int d^{3}r\,
\vec{S}_{JJ1}^M\left(q\vec{r}\right)
\cdot\vec{\mathcal{J}}\left(\vec{r}\right).
\label{eq:multipoles}
\end{align}
The functions appearing here are defined as
\begin{align}
S_{JM}\left(q\vec{r}\right)
&\equiv
j_{J}\left(qr\right)Y_{JM}\left(\hat{r}\right),\\
\vec{S}_{JL1}^M\left(q\vec{r}\right)
&\equiv
j_{L}\left(qr\right)
\vec{Y}_{JL1}^M\left(\hat{r}\right).
\end{align}
In these expressions, $\mathcal{J}_{0}(\vec r)$ and $\vec {\mathcal{J}}(\vec r)$ represent the
nuclear charge and current densities, respectively, and in practical calculations they are typically obtained using the impulse approximation, as will be presented in Subsec. \ref{subsec:multipole_impolse}. The functions $j_{J}$ are
spherical Bessel functions, while $Y_{JM}$ and
$\vec{Y}_{JL1}^{M}$ denote ordinary and vector spherical harmonics.

The expression for $\Theta$ in Eq.~\eqref{eq:multipole_exp} describes the Standard Model $V-A$ structure of the weak interaction with plane-wave
leptonic kinematic factors.
Nuclear-structure-dependent electrostatic effects extend
this shape factor beyond the general correction to the
Fermi function given in Eq.~\eqref{eq:electrostatic_Fermi}.
The Coulomb interaction of the emitted charged lepton
with the nuclear charge distribution modifies both the
leptonic current and the momentum transfer entering the nuclear multipoles.
At first order in the Coulomb interaction, expanding these
contributions and retaining the leading long-wavelength
terms yields corrections that can be expressed through the same
$\hat{C}_J$, $\hat{L}_J$, $\hat{E}_J$, and $\hat{M}_J$
matrix elements, with modified leptonic kinematic
coefficients~[\cite{benatar2026leading}].
The resulting Standard Model shape factor therefore
retains the nuclear multipole basis while acquiring
additional energy and angular dependence.

Beyond the Standard Model, additional scalar, pseudoscalar, and tensor interactions involve different contractions of the leptonic and nuclear currents and therefore require a further extension of the shape factor.
In particular, a tensor current $\mathcal{J}_{\mu\nu}$ cannot be inserted directly into the multipole definitions above. Because this current is antisymmetric, its independent components can instead be organized into two three-vector objects [\cite{glick2023multipole}],
\begin{equation}
\left[\vec{\mathcal{J}}^{\,T}\right]_i
\equiv -\frac{i}{\sqrt{2}}\epsilon_{ijk}\mathcal{J}_{jk},
\qquad
\left[\vec{\mathcal{J}}^{\,T'}\right]_i
\equiv \sqrt{2}\mathcal{J}_{0i},
\label{eq:tensor_decomposition}
\end{equation}
where $i,j,k$ label Cartesian spatial components,  $\epsilon_{ijk}$ is the three-dimensional Levi-Civita symbol, and repeated spatial indices are summed.
The two objects transform under parity as an axial vector and a polar vector, respectively. Substituting each into the definitions of $\hat{L}_J$, $\hat{E}_J$, and $\hat{M}_J$ gives the corresponding tensor multipoles.
Scalar and pseudoscalar nuclear densities can enter through the Coulomb-type multipoles $\hat{C}_J^S$ and $\hat{C}_J^P$. Here, each multipole carries the label of the nuclear current entering its definition. The superscripts $S$, $P$, $V$, and $A$ denote scalar, pseudoscalar, vector, and axial-vector currents, respectively, while $T$ and $T'$ refer to the two objects of the tensor current defined in Eq.~\eqref{eq:tensor_decomposition}.

The resulting shape factor contains the corresponding nuclear multipole matrix elements, with leptonic kinematic factors appropriate to each interaction and interference term. These factors include leptonic structures already present in Eq.~\eqref{eq:multipole_exp}, with modified coefficients and signs, as well as additional terms, such as $m_e/E_e$, where $m_e$ denotes the electron mass, and $E_e$ is the electron energy.

At leading order in the nonrelativistic impulse approximation, the spatial tensor and axial-vector currents are related by
$\vec{\mathcal{J}}^{\,T}
\simeq -\frac{i}{\sqrt{2}}\frac{g_T}{g_A}
\vec{\mathcal{J}}^{\,A}$,
while the scalar and vector charge densities satisfy
$\mathcal{J}^{S}\simeq \frac{g_S}{g_V}\mathcal{J}_0^{V}$.
The mixed time--space tensor current $\vec{\mathcal{J}}^{\,T'}$ first appears at subleading order in the inverse-nucleon-mass expansion.
The leading nonvanishing pseudoscalar Coulomb multipole satisfies
$\hat{C}_J^P\simeq \frac{q}{2m_N}\frac{g_P}{g_A}\hat{L}_J^A$, where the axial longitudinal multipole is also evaluated at leading order. Here $g_V$, $g_A$, $g_S$, $g_T$, and $g_P$ are the vector, axial-vector, scalar, tensor, and pseudoscalar charges of the nucleon, respectively (for the values of these charges, see Sec. 2.2 of \cite{gonzalez2019new}).
The kinematic suppression factor $q/(2m_N)$ in the pseudoscalar contribution is partly offset by the large nucleon pseudoscalar charge $g_P$, associated with the pion pole.
Thus, the leading scalar, pseudoscalar, and tensor multipoles can be expressed in terms of the corresponding Standard Model nuclear matrix elements, allowing the common nuclear matrix elements to be factored out and the dependence on the BSM couplings to be isolated.

Precision $\beta$ decay experiments can then probe BSM couplings through deviations of $\Theta$ from its Standard Model behavior. For illustration, consider an allowed transition from an unpolarized nucleus at leading order in the recoil expansion, neglecting recoil-order contributions including pseudoscalar terms\footnote{The pseudoscalar terms contribute to the Gamow--Teller transition and can be found in~\cite{glick2023multipole}.}. The shape factor can then be
parameterized as
\begin{equation}
\Theta \propto
1+a\,\vec{\beta}\!\cdot\!\hat{\nu}
+b\,\frac{m_e}{E_e},
\end{equation}
where $a$ characterizes the electron--neutrino angular correlation, and $b$ describes the Fierz interference contribution to the electron spectrum. 
Introducing the common normalization $\xi$, these
coefficients, including the Coulomb contribution to $a$,
can be written as~[\cite{lee1956question, JACKSON1957coulomb}]:
\begin{align}
\xi ={}&
|M_F|^2
\left(
|C_V|^2+|C_V'|^2+|C_S|^2+|C_S'|^2
\right)
+
|M_{GT}|^2
\left(
|C_A|^2+|C_A'|^2+|C_T|^2+|C_T'|^2
\right),
\\
a\,\xi ={}&
|M_F|^2
\left(
|C_V|^2+|C_V'|^2-|C_S|^2-|C_S'|^2
\right)
-
\frac{1}{3}|M_{GT}|^2
\left(
|C_A|^2+|C_A'|^2-|C_T|^2-|C_T'|^2
\right)
\nonumber\\
&\mp
\frac{2\alpha Z_f m_e}{k_e}
\operatorname{Im}\left[
|M_F|^2
\left(C_V C_S^*+C_V' C_S'^*\right)
-\frac{1}{3}|M_{GT}|^2
\left(C_A C_T^*+C_A' C_T'^*\right)
\right],
\\
b\,\xi ={}&
\pm 2\gamma_0\,
\operatorname{Re}\left[
|M_F|^2
\left(C_V C_S^*+C_V' C_S'^*\right)
+
|M_{GT}|^2
\left(C_A C_T^*+C_A' C_T'^*\right)
\right],
\end{align}
%
thus, one obtains $a$ and $b$ by $a=a\, \xi/\xi$ and $b=b\,\xi/\xi$. Here the upper (lower) sign corresponds to $\beta^-$ ($\beta^+$) decay,
and $\gamma_0 = \sqrt{1-(\alpha Z_f)^2}$ corrects for relativistic Coulomb effects.
The coefficients denoted by $C$ are effective nucleon-level
couplings in the Lee--Yang parametrization, with the definitions
and sign conventions of~[\cite{gonzalez2019new}].
The unprimed and primed couplings multiply parity-conserving
and parity-violating terms, respectively.
The nucleon charges, $G_F$ and $V_{ud}$ are absorbed into these couplings,
while $M_F$ and $M_{GT}$ are the Fermi and Gamow-Teller nuclear matrix elements, defined without these charges.

The Fierz coefficient $b$ and the Coulomb term in $a$ probe the real and imaginary parts, respectively, of scalar--vector and tensor--axial interference\footnote{%
Within the dimension-six SMEFT, assuming no cancellations among independent contributions, electric dipole moment searches impose tighter indirect bounds on imaginary couplings~[\cite{VosWT2015,gonzalez2019new}].%
}. 
These terms provide linear sensitivity to scalar and tensor couplings, whereas the remaining scalar and tensor corrections to $a$ begin at quadratic order.
In the Standard Model, only vector and axial-vector interactions contribute, with purely left-handed neutrino couplings. 
At this order, a \textbf{pure Fermi} transition ($M_{GT}=0$) therefore has 
 \begin{equation}
a_{\mathrm{SM}}=1,
\qquad
b_{\mathrm{SM}}=0.
\end{equation}
 While a \textbf{pure Gamow--Teller} transition ($M_F=0$) has
\begin{equation}
a_{\mathrm{SM}}=-\frac{1}{3},
\qquad
b_{\mathrm{SM}}=0.
\end{equation}
Measurements are currently targeting deviations from these predictions at the $10^{-3}$ level to probe possible scalar- or tensor-mediated BSM effects [\cite{King2026witepaper1}].
At this precision, however, nuclear-structure contributions predicted within
the Standard Model can no longer be neglected. 
Recoil, spectral-shape, and nuclear-structure-dependent electrostatic effects first enter at next-to-leading order (NLO) in several small parameters. The most relevant parameters are summarized in Table~\ref{tab:expansion_parameters}. These higher-order nuclear effects modify the experimentally inferred values
of both the angular-correlation coefficient and the Fierz term. If they are
not calculated and incorporated consistently, their contributions may either
be misinterpreted as evidence for BSM interactions or conceal genuine
deviations from the Standard Model.

\begin{table}[htbp]
\centering
\begin{tabular}{ccl}
\hline\hline
\textbf{Symbol} & \textbf{Definition} & \textbf{Typical size (endpoint $E_0 \sim 10$~MeV)} \\ [0.5ex] 
\hline
$\epsilon_{\text{NR}}$ & $\dfrac{p_F}{m_N}$ & $\sim 0.2$ (non-relativistic nucleons) \\ [2ex]
$\epsilon_{qr}$ & $qR$ & $\sim 0.05\,A^{1/3}$ (leptonic-wave expansion) \\ [1ex]
$\epsilon_{\text{recoil}}$ & $\dfrac{q}{m_N}$ & $\sim 0.01$ (nucleon recoil) \\ [2ex]
$\epsilon_M$ & $\dfrac{\Delta M}{M_{\min}}$ & $\sim 0.01\,A^{-1}$ (nucleus recoil) \\ [1.5ex]
$\epsilon_{\mathrm{c}}$ & $\alpha Z_f$ & $\sim 0.007 Z_f$  
(Coulomb) \\ [1ex]
\hline\hline
\end{tabular}
\caption{Most relevant small expansion parameters in $\beta$-decay. Here, $p_{\mathrm{F}}$ denotes a typical nuclear Fermi momentum, $m_N$ is the nucleon mass, \(q\) is the
momentum transfer, \(R\) is the nuclear radius, \(A\) is the nuclear
mass number, $\Delta M = M_i - M_f$, and $M_{\min} = \min(M_i, M_f)$, where $M_i$ ($M_f$) is the mass of the initial (final) nuclei. $Z_f$ denotes the charge number of the final nucleus. Because each parameter can be different for a given transition, the ordering of corrections must be judged case-by-case. The estimates involving $q$ correspond to a representative endpoint energy
of 10 MeV.}
\label{tab:expansion_parameters}
\end{table}

\subsubsection{Shape factor and its recoil corrections for allowed $\beta$ decays} \label{sec:shape_allowed}

In the Standard Model, for a \textbf{pure Fermi} transition ($a_\mathrm{SM}=1$), the complete shape factor, including
next-to-leading-order (NLO) shape and recoil effects, can be expressed as~[\cite{glick2022formalism}]:
\begin{equation}
\Theta\left(q,\vec{\beta}\cdot\hat{\nu}\right)
=
\left(1+\delta_{1}\right)
\left[
1+a_\mathrm{SM}
\left(1+\tilde{\delta}_{a}\right)
\vec{\beta}\cdot\hat{\nu}
+(b_\mathrm{SM}+\delta_{b})\frac{m_{e}}{E_{e}}
\right]
\left|\braket{\Vert\hat{C}_{0}^{V}\Vert}\right|^{2}
\label{eq:Fermi}
\end{equation}
The reduced matrix element $\braket{\Vert\hat{C}_{0}^{V}\Vert}$ is associated with the Fermi operator at the leading order (the forms of the multipole operators in the impulse approximation will be given in the next subsection). Since the measurements are looking for any deviations from the Standard Model values of $a$ and $b$, the higher-order corrections $\delta_{1}$,
$\tilde{\delta}_{a}$, and $\delta_{b}$ need to be accounted precisely to avoid confusing them with BSM physics. The terms $\delta_{1}$,
$\tilde{\delta}_{a}$, and $\delta_{b}$ collect contributions from subleading
multipoles and therefore encode NLO recoil and shape effects. They are given as:
\begin{equation}
\begin{split}
\delta_{1}
&=
-2
\frac{E_{0}}{q}
\mathfrak{Re}\left[
\frac{\braket{\Vert\hat{L}_{0}^{V}/q\Vert}}
{\braket{\Vert\hat{C}_{0}^{V}\Vert}}\right]+\frac{E_e-\Delta M}{M_i}
+ \mathcal{O}(\epsilon_{qr}^2 \epsilon_{\mathrm{NR}}^2, \epsilon_\mathrm{recoil}^2, \epsilon_M, \epsilon_{qr}\epsilon_{\mathrm{c}}^2, \epsilon_{qr}^2\epsilon_{\mathrm{c}}),
\\
\tilde{\delta}_{a}
&=
\mathcal{O}(\epsilon_{qr}^2 \epsilon_{\mathrm{NR}}^2, \epsilon_\mathrm{recoil}^2, \epsilon_M, \epsilon_{qr}\epsilon_{\mathrm{c}}^2, \epsilon_{qr}^2\epsilon_{\mathrm{c}}),
\\
\delta_{b}
&=
2
\frac{m_{e}}{q}
\mathfrak{Re}\left[
\frac{\braket{\Vert\hat{L}_{0}^{V}/q\Vert}}
{\braket{\Vert\hat{C}_{0}^{V}\Vert}} \right]
 + \mathcal{O}(\frac{m_e}{E_e}\epsilon_{qr}^2 \epsilon_{\mathrm{NR}}^2,\frac{m_e}{E_e} \epsilon_\mathrm{recoil}^2, \frac{m_e}{E_e}\epsilon_{qr}\epsilon_{\mathrm{c}}^2, \frac{m_e}{E_e}\epsilon_{qr}^2\epsilon_{\mathrm{c}}),
\end{split}
\label{eq:Fermi_corrections}
\end{equation}
Here the arguments of $\mathcal{O}$ show the order of magnitude of the next-order terms which are not presented, as defined in table~\ref{tab:expansion_parameters}. Note that $\epsilon_M$ represents corrections which arise due to the recoil of the nucleus (Eq.~\eqref{eq:recoil_nucleus}). They are presented here although they do not depend on the nuclear structure, so one will be aware that they change those observables and will take them into account for very light nuclei (full terms can be found in~\cite{glick2022formalism}). 

In the Standard Model, for a \textbf{pure Gamow--Teller} transition ($a_\mathrm{SM}=-1/3$), the complete shape factor, including the
leading recoil and spectral-shape corrections, can be written in the compact
form~[\cite{glick2022formalism}]
\begin{equation}
\Theta\left(q,\vec{\beta}\cdot\hat{\nu}\right)
=
3\left(1+\delta_{1}\right)
\left[
1+a_\mathrm{SM}
\left(1+\tilde{\delta}_{a}\right)
\vec{\beta}\cdot\hat{\nu}
+(b_\mathrm{SM}+\delta_{b})\frac{m_{e}}{E_{e}}
\right]
\left|\braket{\Vert\hat{L}_{1}^{A}\Vert}\right|^{2}.
\label{eq:GT}
\end{equation}
The reduced matrix element
$\braket{\Vert\hat{L}_{1}^{A}\Vert}$ at the leading order is associated with the
Gamow--Teller operator. The NLO corrections $\delta_{1}$,
$\tilde{\delta}_{a}$, and $\delta_{b}$ in this case are given by:
\begin{equation}
\begin{split}
\delta_{1}
&=
\frac{2}{3}\mathfrak{Re}\left[
-E_{0}
\frac{\braket{\Vert\hat{C}_{1}^{A}/q\Vert}}
{\braket{\Vert\hat{L}_{1}^{A}\Vert}}
\pm\sqrt{2}\left(E_{0}-2E_{e}\right)
\frac{\braket{\Vert\hat{M}_{1}^{V}/q\Vert}}
{\braket{\Vert\hat{L}_{1}^{A}\Vert}}
+\sqrt{2}
\frac{\braket{\Vert\hat{E}_{1}^{A'}\Vert}}
{\braket{\Vert\hat{L}_{1}^{A}\Vert}}
\right] + \mathcal{O}(\epsilon_{qr}^2 \epsilon_{\mathrm{NR}}^2, \epsilon_\mathrm{recoil}^2, \epsilon_M, \epsilon_{qr}\epsilon_{\mathrm{c}}^2, \epsilon_{qr}^2\epsilon_{\mathrm{c}}),
\\
\tilde{\delta}_{a}
&=
\frac{4}{3}\mathfrak{Re}\left[
2E_{0}
\frac{\braket{\Vert\hat{C}_{1}^{A}/q\Vert}}
{\braket{\Vert\hat{L}_{1}^{A}\Vert}}
\pm\sqrt{2}\left(E_{0}-2E_{e}\right)
\frac{\braket{\Vert\hat{M}_{1}^{V}/q\Vert}}
{\braket{\Vert\hat{L}_{1}^{A}\Vert}}
+\sqrt{2}
\frac{\braket{\Vert\hat{E}_{1}^{A'}\Vert}}
{\braket{\Vert\hat{L}_{1}^{A}\Vert}}
\right] + \mathcal{O}(\epsilon_{qr}^2 \epsilon_{\mathrm{NR}}^2, \epsilon_\mathrm{recoil}^2, \epsilon_M, \epsilon_{qr}\epsilon_{\mathrm{c}}^2, \epsilon_{qr}^2\epsilon_{\mathrm{c}}),
\\
\delta_{b}
&=
\frac{2}{3}m_{e}\mathfrak{Re}\left[
\frac{\braket{\Vert\hat{C}_{1}^{A}/q\Vert}}
{\braket{\Vert\hat{L}_{1}^{A}\Vert}}
\pm\sqrt{2}
\frac{\braket{\Vert\hat{M}_{1}^{V}/q\Vert}}
{\braket{\Vert\hat{L}_{1}^{A}\Vert}}
\right] + \mathcal{O}(\frac{m_e}{E_e}\epsilon_{qr}^2 \epsilon_{\mathrm{NR}}^2,\frac{m_e}{E_e} \epsilon_\mathrm{recoil}^2, \frac{m_e}{E_e}\epsilon_{qr}\epsilon_{\mathrm{c}}^2, \frac{m_e}{E_e}\epsilon_{qr}^2\epsilon_{\mathrm{c}}).
\end{split}
\label{eq:GT_corrections}
\end{equation}
%
Here, $\frac{\hat{C}_{1}^{A}}{\hat{L}_{1}^{A}},\frac{\hat{M}_{1}^{V}}{\hat{L}_{1}^{A}}\propto\epsilon_{\text{recoil}},\epsilon_{qr}\epsilon_{\text{NR}}$, while $\frac{\hat{E}_{1}^{A'}}{\hat{L}_{1}^{A}} \propto \epsilon_{qr}^{2}$ comes from the residual part of the electric multipole that is not fixed by the longitudinal operator:
\begin{equation}
\hat{E}_{J}
=
\sqrt{\frac{J+1}{J}}\hat{L}_{J}
-i\sqrt{\frac{2J+1}{J}}
\int d^{3}r\,
j_{J+1}\left(qr\right)
\vec{Y}_{J,J+1,1}\left(\hat r\right)
\cdot\vec{\mathcal{J}}(\vec r)
\equiv
\sqrt{\frac{J+1}{J}}\hat{L}_{J}
+\hat{E}_{J}'.
\end{equation}

For the pure Fermi and Gamow--Teller transitions considered
here, the additional electrostatic contributions to the
shape factor vanish at leading order in the long-wavelength
expansion~[\cite{benatar2026leading}].
The leading nuclear-structure-dependent electrostatic
correction is therefore fully included by the replacement
in Eq.~\eqref{eq:electrostatic_Fermi}.
When combined with the recoil and shape corrections above,
$\Delta F^\mp_{\mathrm{EC}}$ is retained only as a factor
multiplying the leading-order shape factor.
Products with subleading contributions are neglected.

More generally, transitions between states with angular momenta $J_{\mathrm{i}}$
and $J_{\mathrm{f}}$ receive contributions from every integer $J$ allowed by  the selection rule
\begin{equation}
\left|J_{\mathrm{i}}-J_{\mathrm{f}}\right|
\leq J\leq
J_{\mathrm{i}}+J_{\mathrm{f}}.
\end{equation}
Consequently, transitions involving other angular-momentum combinations
require the corresponding higher-rank multipoles. Those will not be presented here, but can be found in~\cite{glick2022formalism}.

\subsubsection{Impulse approximation forms of multipole operators \label{subsec:multipole_impolse}}

In practical calculations, the multipole operators are evaluated in
the impulse approximation. Their one-body axial and vector forms are
constructed following~\cite{Donnelly:1979ezn} and can be found in~\cite{glick2022formalism}:
\begin{equation}
\begin{split}
\hat{C}_{J}^{A}(q)
&=
\frac{i}{m_{N}}
\sum_{j=1}^{A}\tau_{j}^{\pm}
\left[
g_{A}S_{J}(q\vec r_{j})
\vec\sigma_{j}\cdot\vec\nabla
+\frac{1}{2}
\left(
g_{A}
-\frac{E_{0}\pm\Delta E_{c}}{2m_{N}}\tilde g_{P}
\right)
\vec\nabla S_{J}(q\vec r_{j})
\cdot\vec\sigma_{j}
\right],
\\
\hat{L}_{J}^{A}(q)
&=
\frac{i}{q}
\left(
g_{A}+\frac{q^{2}}{4m_{N}^{2}}\tilde g_{P}
\right)
\sum_{j=1}^{A}\tau_{j}^{\pm}
\vec\nabla S_{J}(q\vec r_{j})
\cdot\vec\sigma_{j},
\\
\hat{E}_{J}^{A}(q)
&=
\frac{1}{q}
\left(
g_{A}+\frac{q^{2}}{4m_{N}^{2}}\tilde g_{P}
\right)
\sum_{j=1}^{A}\tau_{j}^{\pm}
\left[
\vec\nabla\times
\vec S_{JJ1}(q\vec r_{j})
\right]\cdot\vec\sigma_{j},
\\
\hat{M}_{J}^{A}(q)
&=
\left(
g_{A}+\frac{q^{2}}{4m_{N}^{2}}\tilde g_{P}
\right)
\sum_{j=1}^{A}\tau_{j}^{\pm}
\vec S_{JJ1}(q\vec r_{j})\cdot\vec\sigma_{j},
\\
\hat{C}_{J}^{V}(q)&=g_V \sum_{j=1}^A \tau_j^{\pm} S_J (q\vec{r}_j) ,
\\
\hat{L}_{J}^{V}(q)&= - \frac{q}{2m_N} g_V \sum_{j=1}^A \tau_j^\pm \left[ S_J(q\vec{r}_j)-\frac{2}{q^2} [\vec\nabla S_J(q\vec{r}_j)]\cdot\vec\nabla \right]  ,
\\
\hat{E}_{J}^{V}(q)&= \frac{q}{m_N} \sum_{j=1}^A \tau_j^\pm \left[ -i\frac{g_V}{q^2} [\vec \nabla \times \vec S_{JJ1}(q\vec{r}_j)]\cdot \vec\nabla + \frac{\mu}{2} \vec{S}_{JJ1}(q\vec r) \cdot \vec \sigma_j\right],
\\
\hat{M}_{J}^{V}(q)
&=
-\frac{i}{m_{N}}
\sum_{j=1}^{A}\tau_{j}^{\pm}
\left\{
g_{V}\vec S_{JJ1}(q\vec r_{j})\cdot\vec\nabla
+\frac{i}{2}\mu
\left[
\vec\nabla\times
\vec S_{JJ1}(q\vec r_{j})
\right]\cdot\vec\sigma_{j}
\right\}.
\end{split}
\label{eq:multipoles1b}
\end{equation}
Here, we omitted the angular momentum projection index $M$ for simplicity, as these operators appear in Eq. \eqref{eq:multipole_exp} and subsequent equations only inside reduced matrix elements. The Coulomb-energy shift $\Delta E_{c}$ accounts for additional
electromagnetic effects and is included following
~\cite{behrens1982electron} as was discussed in Sec. \ref{sec:kinematics}. The index $j$ labels the
nucleons, and all sums extend from $1$ to the mass number $A$. The operators
$\vec\sigma_{j}/2$ and $\tau_{j}/2$ are the spin and isospin operators of the
$j$th nucleon. The quantities $g_{A}$, $g_{V}$, $\tilde g_{P}$, and $\mu$
denote the axial, vector, induced-pseudoscalar, and magnetic form factors,
respectively. Their values are $g_V (0) \simeq1$, $g_A(0)\simeq 1.27$, $\mu\simeq4.706$ [\cite{PDG2020}] and 
\begin{equation}
\tilde g_{P}(q^{2})
\simeq
-\frac{(2m_{N})^{2}}
{m_{\pi}^{2}-q^{2}}\,g_{A}(0).
\end{equation}
The spherical functions $S_{J}\left(q\vec{r}\right) \equiv
j_{J}\left(qr\right)Y_{J}\left(\hat{r}\right)$ and $\vec{S}_{JL1}\left(q\vec{r}\right) \equiv
j_{L}\left(qr\right)
\vec{Y}_{JL1}\left(\hat{r}\right)$ appear due to the plane-wave expansion of the lepton wavefunctions discussed in Sec. \ref{sec:kinematics}. 

Derivations of Fermi, Gamow-Teller and weak magnetism operators from these forms of the multipole operators in the long-wavelength limit are given in the Appendix.

\subsection{Connection to Holstein's form factors}
Different representations of the expressions for calculating recoil-order corrections have been used in the literature. Most notably, forms of the recoil-order operators by \cite{Holstein:1974zf} have been widely used. Here we show the connection between the Donnelly-Walecka's and Holstein's representations.

Holstein's form-factor framework provides an alternative, complementary description of allowed $\beta$ decay. It is commonly used in analyses of experimental data and captures the leading long-wavelength nuclear-structure effects through a small set of reduced form factors. The leading Fermi and Gamow--Teller contributions are represented by $a_V$ 
and $c_A$, respectively. The principal recoil-order Standard-Model terms are encoded in the weak-magnetism form factor $b_{\mathrm{WM}}$, the induced-tensor form factor $d_\text{IT}$, and the second-forbidden vector, $f_V$, and axial form factors $j_K^A$.
The definitions below use the phase convention specified in the footnote.\footnote{In \cite{Holstein:1974zf}, the angular momenta are coupled so that the final-state angular momentum is obtained by coupling the spherical tensor to the initial state. Here, the initial state is coupled to the spherical tensor to produce the final state. Consequently, the expressions in Eq.~\eqref{eq:Holstein} differ from those of Ref.~\cite{Holstein:1974zf} by a factor $(-1)^{J_f-J_i}/\sqrt{2J_i+1}$. In most practical calculations, the recoil order terms are reported as fractions with respect to $c$ or $a$, so this factor cancels out.}
\begin{eqnarray}
a_V &=& \frac{(-1)^{J_f - J_i}}{\sqrt{2J_i+1}} g_V \braket{J_f \Vert \sum_{j=1}^A \tau_j^{\pm} \Vert J_i}, \nonumber \\
b_{\mathrm{WM}}
&=&
A\frac{(-)^{(J_f-J_i)}}{\sqrt{2J_i+1}}
\Bigg[
g_V
\braket{J_f \Vert \sum_{i=1}^A \tau^{\pm}_i l_i \Vert J_i}
+
g_M
\braket{J_f \Vert \sum_{i=1}^A \tau^{\pm}_i \sigma_i \Vert J_i}
\Bigg],
\nonumber\\
c_A
&=&
\frac{(-)^{(J_f-J_i)}}{\sqrt{2J_i+1}}\,
g_A
\braket{J_f \Vert \sum_{i=1}^A \tau^{\pm}_i \sigma_i \Vert J_i}
=
\frac{(-)^{(J_f-J_i)}}{\sqrt{2J_i+1}}\,g_A M_{GT},
\nonumber\\
d_\mathrm{IT}
&=&
A\frac{(-)^{(J_f-J_i)}}{\sqrt{2J_i+1}}\,
g_A
\braket{J_f \Vert
\sum_{i=1}^A \tau^{\pm}_i \sqrt{2}[l_i \times \sigma_i]^1
\Vert J_i},
\nonumber\\
f_V &=& \frac{(-1)^{J_f - J_i}}{\sqrt{2J_i+1}}  \frac{A m_N E_0}{\sqrt{6}} g_V \braket{J_f \Vert \sum_{j=1}^A \tau_j^{\pm}  \hat Q_j \Vert J_i}
\nonumber \\
j_K^A
&=&
-\frac{2}{3}A^2m_N^2
\frac{(-)^{(J_f-J_i)}}{\sqrt{2J_i+1}}\,
g_A
\braket{J_f \Vert
\sum_{j=1}^A \tau^{\pm}_j [\hat Q_j \times \sigma_j]^K
\Vert J_i},
\qquad K=2,3.
\label{eq:Holstein}
\end{eqnarray}
Here, $l_i$ is the single-particle orbital angular-momentum operator, and
$\hat Q_i=\sqrt{16\pi/5}\,r_i^2Y_{2\mu}(\hat r_i)$ is the quadrupole operator. Thus, $d_\mathrm{IT}$ characterizes the axial spin-orbit recoil contribution, $b_{\mathrm{WM}}$ describes the vector magnetic response, $f_V$ describes the quadrupole response, and $j_K^A$ is built from quadrupole--spin matrix elements. Contributions from $f_V$ and $j_K^A$ arise only for transitions whose angular-momentum selection rules permit a rank-2 or rank-$K$ operator. Two other form factors from \cite{Holstein:1974zf}, namely induced scalar $e_\mathrm{IS}$ and second forbidden vector $g$ are related to $a_V$ and $f_V$, respectively. Assuming the weak
vector current to be divergenceless (CVC), for transitions between non-isobaric analog states the following relations hold [\cite{Tribble1975}]:
\begin{align}
    e_\mathrm{IS}&=-\frac{2Am_N E_0}{q^2}a_V, \nonumber \\
    g&=-\sqrt{\frac{8}{3}}\frac{Am_N}{E_0}f_V. \nonumber
\end{align}

The Donnelly--Walecka multipole formalism and Holstein's parametrization describe the same underlying physics in different ways. The former organizes the full response of the nuclear current into multipoles, whereas the latter gives their leading long-wavelength limits. When $q/m_N\ll 1$ and $qR\ll 1$, expanding the spherical Bessel functions in the multipole operators in powers of $q/m_N$ and $qR$ yields the leading one-body operators that appear in Eq.~\eqref{eq:Holstein}.

At leading order in $qR$, the rank-zero and rank-one matrix elements of the one-body impulse-approximation operators in Eq.~\eqref{eq:multipoles1b} can be expressed in terms of Holstein's form factors as
\begin{equation}
\label{eq:mapping}
\begin{split}
\braket{J_f\Vert \hat{C}_0^V \Vert J_i}
&\simeq
\frac{\sqrt{2J_i+1}}{(-)^{(J_f-J_i)}}
\frac{a_{\mathrm{V}}}{\sqrt{4\pi}}. \\
\braket{J_f\Vert \hat{M}_1^V/q \Vert J_i}
&\simeq
i\,
\frac{\sqrt{2J_i+1}}{(-)^{(J_f-J_i)}}
\frac{b_{\mathrm{WM}}}{2\sqrt{6\pi}A m_N}. \\
\braket{J_f\Vert \hat{L}_1^A \Vert J_i}
&\simeq
i\,
\frac{\sqrt{2J_i+1}}{(-)^{(J_f-J_i)}}
\frac{c_A}{2\sqrt{3\pi}},
\\
\braket{J_f\Vert \hat{C}_1^A/q \Vert J_i}
&\simeq
i\,
\frac{\sqrt{2J_i+1}}{(-)^{(J_f-J_i)}}
\frac{d_\mathrm{IT}}{2\sqrt{3\pi}A m_N}.
\end{split}
\end{equation}
Note that the axial Coulomb multipole $\hat{C}_1^A$ requires a more careful consideration, as it includes the Coulomb displacement energy correction $\Delta E_c/(2m_N)$, which is missing in Holstein's parametrization although it may be numerically significant in some transitions; see, for example,~\cite{glick2026ab1}.
In obtaining Eq.~\eqref{eq:mapping}, the complex couplings in the multipole operators of Eq.~\eqref{eq:multipoles1b} have been replaced by $g_A(q^2)$ in the strict long-wavelength limit. This approximation is appropriate for $\hat{L}_1^A$, $\hat{E}_1^A$, and $\hat{M}_1^V$, for which the induced-pseudoscalar contribution is suppressed by $q^2/(2m_N)^2$.
A detailed derivation for the weak magnetism $b_\text{WM}$ form factor is available in the Appendix.

Similarly, for the rank-2 vector form factor $f_V$ one obtains
\begin{equation}
   \braket{J_f\Vert \hat{C}_2^V /q^2 \Vert J_i}
\simeq \frac{\sqrt{2J_i+1}}{(-)^{(J_f-J_i)}}\frac{1}{2\sqrt{30\pi}} \frac{f_V}{Am_N E_0}.
\end{equation}
A corresponding long-wavelength expansion connects higher-rank axial multipoles with the quadrupole--spin operators that define the form factors $j_K^A$. In the conventions of Eq.~\eqref{eq:Holstein}, the relations are
\begin{equation}
\begin{split}
\braket{J_f\Vert \hat{M}_2^A(q)/q^2 \Vert J_i}
&\simeq
-
\frac{\sqrt{2J_i+1}}{(-)^{J_f-J_i}}
\frac{j_2^A}{40\sqrt{5\pi}\,A^2m_N^2},
\\[1em]
\braket{J_f\Vert \hat{L}_3^A/q^2 \Vert J_i}
&\simeq
-i
\frac{\sqrt{2J_i+1}}{(-)^{J_f-J_i}}
\frac{3}{8}
\frac{j_3^A}{\sqrt{105\pi}\,A^2m_N^2}.
\end{split}
\end{equation}
These contributions are present only when the initial and final nuclear angular momenta allow reduced matrix elements of rank two or rank three, respectively.

\section{Nuclear structure calculations of recoil-order corrections}

This section describes the nuclear-structure framework used to evaluate recoil-order corrections in nuclear $\beta$ decay. We begin by outlining the nuclear many-body problem and the methods used to obtain nuclear states and transition densities. We then provide the formalism needed to calculate the multipole matrix elements in the harmonic oscillator basis.

\subsection{Nuclear Many-Body Methods}

The calculation of recoil-order corrections in beta decay requires accurate descriptions of nuclear structure, which necessitates solving the nuclear many-body problem. Over the past decades, a variety of theoretical approaches have been developed and applied to this challenge, ranging from phenomenological models to increasingly sophisticated \emph{ab initio} techniques. Nuclear many-body methods have proven instrumental in providing essential nuclear structure information for beta decay physics across multiple domains. For precision measurements and the determination of fundamental parameters such as $V_{ud}$, these methods enable the computation of radiative corrections to super-allowed beta transitions~[\cite{Seng:2023cvt,Gennari:2024sbn}]. Additionally, beyond the Standard Model searches utilizing scalar and tensor interactions in the weak sector have relied on nuclear recoil-order corrections computed through many-body approaches [\cite{glick2022nuclear,glick2026ab1,Sargsyan_A8, sargsyan2026role}]. 

The foundation for all nuclear many-body calculations rests upon solving the nuclear Schr\"{o}dinger equation with an appropriate nuclear Hamiltonian. The many-body methods employed in nuclear structure calculations generally fall into two broad categories: coordinate-space methods, which directly manipulate and optimize nuclear wave functions, and configuration-space methods, which expand the nuclear eigenstates as superpositions of many-body basis states~[\cite{Hergert:2020bxy}]. In both approaches, one aims to solve the $A$-body Schr\"{o}dinger equation with a translationally-invariant Hamiltonian of the form
\begin{equation}
    H = \sum_i T_i - T_{CM} + V_\mathrm{Coul} + \sum_{i<j} v_{ij} + \sum_{i<j<k} V_{ijk} + \ldots \, , \label{Eq:intH}
\end{equation}
where $T_i$ denotes the kinetic energy of the nucleon $i$, $T_{CM}$ represents the center-of-mass kinetic energy that must be removed to isolate intrinsic motion, $V_\mathrm{Coul}$ is the Coulomb interaction among protons, and $v_{ij}$ and $V_{ijk}$ are the two- and three-nucleon interactions, respectively. Higher-order many-nucleon forces may exist in principle, but they are expected to be much weaker.

Once a specific model for the Hamiltonian is chosen, one must solve
\begin{equation}
    H\ket{\Psi(J^{\pi};T,T_z)} = E\ket{\Psi(J^{\pi};T,T_z)}\, , \label{Eq:ShrEqn}
\end{equation}
to obtain the nuclear wave function $\Psi$ with the required quantum numbers: total angular momentum $J$, parity $\pi$, isospin $T$, and isospin projection $T_z$, corresponding to energy eigenvalue $E$. Once the nuclear wave functions are established, one can evaluate the weak transition operators, \textit{i.e.}, the charge operator and current operator, to extract observables relevant to beta decay processes.

Various nuclear many-body methods have been employed for calculations of beta decay recoil-order corrections. They can be broadly classified into three levels of sophistication, each with distinct assumptions and ranges of applicability.

\textit{Phenomenological methods} represent the most simplified approach, where the many-body Hamiltonian is constructed from empirical parameters fitted to reproduce known nuclear binding energies, excitation spectra, and other experimental observables. While computationally efficient and historically important for establishing many foundational insights, these approaches sacrifice microscopic foundation for practical tractability.

The \textit{nuclear shell model} constructs the Hamiltonian from two-body matrix elements derived from realistic nucleon-nucleon interactions, typically restricted to a finite model space defined by an inert core and active nucleons distributed among a limited set of single-particle orbitals. The drastic truncation of the model space necessitates the development of an effective interaction that captures the physics of the omitted degrees of freedom. Within the Fock-space formulation, such an effective Hamiltonian is fully specified by two ingredients: a collection of single-particle energies $\varepsilon_i$ that set the reference scale for the valence orbitals relative to an inert core, and a set of two-body interaction matrix elements $V_{ij,kl}$ characterizing the residual inter-particle correlations (under the assumption that the effective interaction is truncated at two-body order):
\begin{equation}
H = \sum_i \varepsilon_i a^{\dagger }_i a_i+ \sum\limits_{ijkl}V_{ij,kl} a^{\dagger }_ia^{\dagger }_j a_l a_k \,.
\end{equation}
There exist two principal routes for determining these Hamiltonian parameters. The first involves deriving them from first principles through microscopic calculations based on realistic bare nucleon-nucleon forces~[\cite{stroberg2019nonempirical,coraggio2024role}]. The second adopts an effective potential wherein the parameters are adjusted to reproduce the known excitation energies and spectra of nuclei within the chosen model space. Despite its empirical nature, the latter approach has demonstrated remarkable success in describing the low-lying nuclear structure across diverse nuclei~[\cite{caurier2005shell}]. 
This approach has been extensively applied to beta-decay calculations, providing excellent agreement with experiment across a wide range of nuclei, though with the systematic limitation that larger model spaces and beyond-two-body forces are not easily incorporated, as well as the absence of a systematic uncertainty quantification.

In recent years, \emph{ab initio} methods have emerged as the leading approach for precision nuclear structure calculations relevant to beta decay. These methods start from fundamental nucleon-nucleon and many-nucleon interactions derived from chiral effective field theory~[\cite{machleidt2011chiral, hammer2019nuclear}], which provides a systematic expansion in powers of a small parameter and connects nuclear forces to the underlying QCD symmetries. The interactions employed in \emph{ab initio} calculations typically depend on the choice of low-energy constants and regularization scheme, introducing a systematic uncertainty that reflects the freedom inherent in effective field theory constructions. Beyond interaction uncertainties, additional theoretical uncertainties arise from the numerical implementation of these interactions within specific many-body frameworks~[\cite{Hergert:2020bxy}]. Despite these limitations, \emph{ab initio} methods provide access to larger model spaces and systematic inclusion of many-body forces, enabling more rigorous theoretical predictions for recoil-order corrections in nuclear beta decay.

\subsection{Multipole operators in the harmonic oscillator basis}

Many nuclear many-body methods, including the nuclear shell model and several \emph{ab initio} approaches, employ a harmonic-oscillator (HO) single-particle basis. We therefore present the multipole operators in this basis, beginning with the second-quantized form of a general one-body operator $\hat O_{JM}$. Its reduced matrix elements are specified between single-particle states labeled by $\ket{n_i,(l_i,\frac{1}{2}),j_i}$
and $\bra{n_f,(l_f,\frac{1}{2}),j_f}$, where the subscripts $i$ and $f$ denote the initial and final states, respectively:
\begin{equation}
    \langle J_f \vert\vert \hat{O}_J \vert\vert J_i \rangle = \frac{1}{\sqrt{2J+1}} \sum_{\substack{n_f l_f j_f \\ n_i l_i j_i}} \langle n_f l_f j_f \vert\vert \hat{O}_J \vert\vert n_i l_i j_i \rangle \langle J_f \vert\vert \left[ \hat{a}^\dagger_{n_f l_f j_f} \times \tilde{a}_{n_i l_i j_i} \right]_J \vert\vert J_i \rangle, 
\end{equation}
where we dropped the nucleon spin $\frac{1}{2}$ label for brevity. The creation and annihilation operators, $\hat a^\dagger$ and $\tilde a$, carry good angular momenta. The second bracket, $\langle J_f \vert\vert [ \hat{a}^\dagger_{n_f l_f j_f} \times \tilde{a}_{n_i l_i j_i} ]_J \vert\vert J_i \rangle$, is the one-body density matrix element, which is independent of the operator shape and can be calculated using the many-body wavefunctions from Eq. \eqref{Eq:ShrEqn}. Hence, to obtain the reduced matrix elements of the multipole operators that enter the formula for the shape factor $\Theta$ in Eq. \eqref{eq:multipole_exp}, we need to evaluate the single-particle matrix elements $\langle n_f l_f j_f \vert\vert \hat{O}_J \vert\vert n_i l_i j_i \rangle $ in the HO basis. For this, we use the impulse approximation forms of the multipole operators in Eq. \eqref{eq:multipoles1b} and notice that there are only four distinct forms of operators that act on the single-particle states, namely, $S_J$, $\vec S_{JL} \cdot \vec \sigma$, $\vec S_{JL} \cdot \vec \nabla/q$ and $S_J \, \vec \sigma \cdot \vec \nabla / q$. This follows from the fact that the gradient of $S_{J}$ obeys the following relation:
\begin{equation}
    \frac{1}{q} \vec{\nabla} S_{JM}(q\vec{r}) = \sqrt{\frac{J}{2J+1}}\vec{S}_{J,J-1,1}^M(q\vec{r}) - \sqrt{\frac{J+1}{2J+1}}\vec{S}_{J,J+1, 1}^M(q\vec{r}).
\end{equation}
Dotted with the Pauli spin vector $\vec\sigma$, it becomes:
\begin{equation}
    \left[ \vec{\nabla} S_{JM}(q\vec{r}) \right] \cdot \vec{\sigma} = q \left[ \sqrt{\frac{J}{2J+1}} \vec{S}_{J,J-1,1}^M(q\vec{r})\cdot\vec{\sigma} - \sqrt{\frac{J+1}{2J+1}} \vec{S}_{J,J+1,1}^M(q\vec{r})\cdot\vec{\sigma} \right].
\end{equation}
While for the cross products, we use the curl identities
\begin{equation}
\left[ \vec{\nabla} \times \vec{S}_{JJ1}^M(q\vec{r}) \right] \cdot \vec{\sigma} = i q \left[ \sqrt{\frac{J+1}{2J+1}} \vec{S}_{J,J-1,1}^M(q\vec{r})\cdot\vec{\sigma} + \sqrt{\frac{J}{2J+1}} \vec{S}_{J,J+1,1}^M(q\vec{r})\cdot\vec{\sigma} \right],
\end{equation}
and
\begin{equation}
    \left[ \vec{\nabla} \times \vec{S}_{JJ1}^M(q\vec{r}) \right] \cdot \vec{\nabla} = i q \left[ \sqrt{\frac{J+1}{2J+1}} \vec{S}_{J,J-1,1}^M(q\vec{r})\cdot\vec{\nabla} + \sqrt{\frac{J}{2J+1}} \vec{S}_{J,J+1,1}^M(q\vec{r})\cdot\vec{\nabla} \right].
\end{equation}

The reduced matrix elements of the four distinct operators can be calculated analytically in the HO basis. In the following equations, we drop the index ``1'' of the vector spherical functions. Following \cite{haxton2008seven}, we factor out the radial parts: 
\begin{align}
\left\langle n_f \left(l_f {\textstyle{1 \over 2}}\right)j_f
\middle\| S_J(qr) \middle\|
n_i \left(l_i {\textstyle{1 \over 2}}\right)j_i\right\rangle
&= {1 \over \sqrt{4 \pi}}(-)^{J+j_i+1/2}
\left[l_f\right]\left[l_i\right]\left[j_f\right]\left[j_i\right]\left[J\right]
 \\
&\times
\left\{ \begin{array}{ccc}
l_f & j_f & {\textstyle {1 \over 2}}  \\
j_i & l_i & J
\end{array} \right\}
\left( \begin{array}{ccc}
l_f & J & l_i \nonumber \\
0 & 0 & 0
\end{array} \right)
\left\langle n_f l_f j_f \middle| j_J(\rho) \middle| n_i l_i j_i \right\rangle
\label{sjbracket} \\
\left\langle n_f \left(l_f {\textstyle{1 \over 2}}\right)j_f
\middle\| \vec{S}_{J L}(qr)\cdot\vec{\bsig} \middle\|
n_i \left(l_i {\textstyle{1 \over 2}}\right)j_i\right\rangle
&= \sqrt{{3 \over 2 \pi}}(-)^{l_f}
\left[l_f\right]\left[l_i\right]\left[j_f\right]\left[j_i\right]\left[L\right]\left[J\right]
 \\
&\times
\left\{ \begin{array}{ccc}
l_f & l_i & L \nonumber \\
{\textstyle {1 \over 2}} & {\textstyle {1 \over 2}} & 1 \nonumber \\
j_f & j_i & J
\end{array} \right\}
\left( \begin{array}{ccc}
l_f & L & l_i \nonumber \\
0 & 0 & 0
\end{array} \right)
\left\langle n_f l_f j_f \middle| j_L(\rho) \middle| n_i l_i j_i \right\rangle
\label{sjlsigmabracket} \nonumber\\
\left\langle n_f \left(l_f {\textstyle{1 \over 2}}\right)j_f
\middle\| \vec{S}_{J L}(qr)\cdot {1 \over q}\vec{\bnab} \middle\|
n_i \left(l_i {\textstyle{1 \over 2}}\right)j_i\right\rangle
&= {1 \over \sqrt{4 \pi}}(-)^{L+j_i+1/2}
\left[l_f\right]\left[l_i\right]\left[j_f\right]\left[j_i\right]\left[L\right]\left[J\right]
\\
&\times
\left\{ \begin{array}{ccc}
l_f & j_f & {\textstyle {1 \over 2}} \\
j_i & l_i & J
\end{array} \right\}\nonumber\\
&\times \Biggl\{
-(l_i+1)^{1/2}\left[l_i+1\right]
\left\{ \begin{array}{ccc}
L & 1 & J \\
l_i & l_f & l_i+1
\end{array} \right\}
\left( \begin{array}{ccc}
l_f & L & l_i+1 \nonumber \\
0 & 0 & 0
\end{array} \right)
\nonumber \\
&\times
\left\langle n_f l_f j_f \middle|
j_L(\rho)\left( {d \over d\rho} - {l_i \over \rho}\right)
\middle| n_i l_i j_i \right\rangle
\nonumber\\
&+ l_i^{1/2}\left[l_i-1\right]
\left\{ \begin{array}{ccc}
L & 1 & J \nonumber \\
l_i & l_f & l_i-1
\end{array} \right\}
\left( \begin{array}{ccc}
l_f & L & l_i-1 \nonumber \\
0 & 0 & 0
\end{array} \right)
\\
&\times
\left\langle n_f l_f j_f \middle|
j_L(\rho)\left( {d \over d\rho} + {l_i+1 \over \rho}\right)
\middle| n_i l_i j_i \right\rangle
\Biggr\}
\label{sjlnablabracket} \nonumber\\
\left\langle n_f \left(l_f {\textstyle{1 \over 2}}\right)j_f
\middle\| S_J(qr)\vec{\bsig}\cdot {1 \over q}\vec{\bnab} \middle\|
n_i \left(l_i {\textstyle{1 \over 2}}\right)j_i\right\rangle
&= {1 \over \sqrt{4 \pi}}(-)^{l_f}
\left[l_f\right]\left[j_f\right]\left[j_i\right]\left[2j_i-l_i\right]\left[J\right] 
\\
&\times
\left\{ \begin{array}{ccc}
l_f & j_f & {\textstyle {1 \over 2}} \nonumber \\
j_i & 2j_i-l_i & J
\end{array} \right\}
\left( \begin{array}{ccc}
l_f & J & 2j_i-l_i \\
0 & 0 & 0
\end{array} \right)
\nonumber\\
&\times \Biggl\{
-\delta_{j_i,l_i+1/2}
\left\langle n_f l_f j_f \middle|
j_J(\rho)\left( {d \over d\rho} - {l_i \over \rho}\right)
\middle| n_i l_i j_i \right\rangle
\nonumber \\
&+\delta_{j_i,l_i-1/2}
\left\langle n_f l_f j_f \middle|
j_J(\rho)\left( {d \over d\rho} + {l_i+1 \over \rho}\right)
\middle| n_i l_i j_i \right\rangle
\Biggr\}. \nonumber
\label{sjnablabracket}
\end{align}
Here $\rho\equiv qr$, $[j]=\sqrt{2j+1}$ and we used Wigner's 3-$j$, 6-$j$ and 9-$j$ symbols for (de)coupling. The radial matrix elements can be evaluated by integrating them with the radial wavefunctions. In the HO basis, these matrix elements can be calculated analytically. The HO radial wavefunctions do not depend on spin, hence, $R_{nlj}\equiv R_{nl}$. The normalized HO radial wave functions are given as
\begin{equation}
R_{nlj}(x) \equiv R_{nl}(x) = \left[ {2 e^z \over b^3 (n-1)! \Gamma(n+l+{\textstyle {1 \over 2}}) z^{l+1}} \right]^{1/2}
 {d^{n-1} \over dz^{n-1}} \left[ z^{n+l-1/2} e^{-z} \right],
\end{equation}
where $z \equiv (x/b)^2$ and $b=\sqrt{\hbar/m_N\Omega}$ is the oscillator length, with $\Omega$ being its frequency.  We follow the notation of \cite{haxton2008seven}
where the radial quantum number is given by $n = (N-l)/2+1$, with the principal quantum number $N$ = 0,1,2,.... Hence, $n$ starts at one. With this, we can evaluate the three distinct radial matrix elements:
\begin{eqnarray}
&&\langle n_f l_f | j_L(\rho) | n_i l_i \rangle = {2^L \over (2L+1)!!} y^{L/2} e^{-y}
\sqrt{(n_i-1)! (n_f-1)! \Gamma(n_f+l_f+{\textstyle {1 \over 2}}) \Gamma(n_i+l_i+{\textstyle {1 \over 2}})}
\sum_{k=0}^{n_i-1} \sum_{k'=0}^{n_f-1}
{(-1)^{k+k'} \over k! k'!} \nonumber \\
&&~~~~ \times~ {1 \over (n_i-1-k)! (n_f-1-k')!}
{\Gamma[{\textstyle {1 \over 2}} (l_i+l_f+L+2k+2k'+3)] \over
\Gamma[l_i+k+{\textstyle {3 \over 2}}] \Gamma[l_f+k'+{\textstyle {3 \over 2}}]} ~{}_1 F_1[{\textstyle {1 \over 2}}(L-l_i-l_f-2k-2k');L+{\textstyle{3 \over 2}};y], \\
&&\langle n_f l_f | j_L(\rho) \left({d \over d\rho} - {l_i \over \rho}\right) | n_i l_i \rangle ={2^{L-1} \over (2L+1)!!} y^{(L-1)/2} e^{-y}
\sqrt{(n_i-1)! (n_f-1)! \Gamma(n_f+l_f+{\textstyle {1 \over 2}}) \Gamma(n_i+l_i+{\textstyle {1 \over 2}})} \nonumber \\
&&~~~~\times ~\sum_{k=0}^{n_i-1} \sum_{k'=0}^{n_f-1}
{(-1)^{k+k'} \over k! k'!} {1 \over (n_i-1-k)! (n_f-1-k')!} {\Gamma[{\textstyle {1 \over 2}}(L+l_i+l_f+2k+2k'+2)] \over
\Gamma[l_i+k+{\textstyle {3 \over 2}}] \Gamma[l_f+k'+{\textstyle {3 \over 2}}]} \nonumber \\
&&~~~~\times \Biggl\{ -{\textstyle{1 \over 2}}(l_i+l_f+L+2k+2k'+2)  \nonumber \\ 
&& ~~~~ \times ~{}_1 F_1[{\textstyle {1 \over 2}} (L-l_i-l_f-2k-2k'-1);
L+{\textstyle {3 \over 2}};y] + (2k) {}~_1 F_1[{\textstyle {1 \over 2}}(L-l_i-l_f-2k-2k'+1);L+{\textstyle {3 \over 2}};y] \Biggr\}, \\
&&\langle n_f l_f | j_L(\rho) \left({d \over d\rho} + {l_i+1 \over \rho}\right) | n_i l_i \rangle ={2^{L-1} \over (2L+1)!!} y^{(L-1)/2} e^{-y}
\sqrt{(n_i-1)! (n_f-1)! \Gamma(n_f+l_f+{\textstyle {1 \over 2}}) \Gamma(n_i+l_i+{\textstyle {1 \over 2}})} \nonumber \\
&&~~~~\times~ \sum_{k=0}^{n_i-1} \sum_{k'=0}^{n_f-1}
{(-1)^{k+k'} \over k! k'!} {1 \over (n_i-1-k)! (n_f-1-k')!} {\Gamma[{\textstyle {1 \over 2}}(L+l_i+l_f+2k+2k'+2)] \over
\Gamma[l_i+k+{\textstyle {3 \over 2}}] \Gamma[l_f+k'+{\textstyle {3 \over 2}}]} \nonumber \\
&&~~~~\times \Biggl\{ -{\textstyle {1 \over 2}}(l_i+l_f+L+2k+2k'+2) \nonumber \\ 
&& ~~~~ \times~{}_1 F_1[{\textstyle {1 \over 2}} (L-l_i-l_f-2k-2k'-1);
L+{\textstyle {3 \over 2}};y] + (2l_i+2k+1) {}~_1 F_1[{\textstyle {1 \over 2}} (L-l_i-l_f-2k-2k'+1);L+{\textstyle {3 \over 2}};y] \Biggr\},
\end{eqnarray}
where $ y = (q b/2)^2$. Here  ${}_1F_1$ is the confluent hypergeometric function and $\Gamma$ is the Euler gamma function. Both of these special functions are available in modern computational libraries. 

With this, we have presented all the ingredients needed to evaluate the nuclear shape factor $\Theta$ that consists of multipole reduced matrix elements. We note that the expressions we obtained in this section do not assume the long-wavelength approximation, so one can use them to evaluate multipole matrix elements at any order. This is, perhaps, one of the largest advantages of the Donnelly--Walecka formalism as it allows one to easily evaluate higher-order terms and assess their contribution for a specific case.

\section{Summary and outlook}

Precision measurements of nuclear $\beta$ decay provide sensitive tests of the charged-current interaction, from determinations of Standard-Model parameters such as $V_{ud}$ to searches for nonstandard scalar, pseudoscalar, tensor, and right-handed interactions. Interpreting these measurements requires theoretical predictions that account for effects beyond the leading allowed approximation. This chapter has focused on recoil-order corrections, which arise from subleading terms in the nuclear weak current and can modify $\beta$-decay spectra, angular correlations, and asymmetries. Though generally small, their contributions---and uncertainties---can be comparable to the effects sought in precision searches for new physics.

We have outlined the relevant $\beta$-decay kinematics and the multipole expansion of the shape factor, identifying the recoil-order operators and form factors that contribute to allowed transitions. We have also presented the impulse-approximation forms of the multipole operators and described how their matrix elements can be evaluated using nuclear many-body states and transition densities. In particular, expressing the operators in a harmonic-oscillator basis provides ingredients applicable to a range of nuclear-structure methods. The Donnelly--Walecka formalism does not require the long-wavelength approximation, making it possible to include higher-order multipoles and assess their importance for a given transition. Radiative corrections are treated in the accompanying chapter; here, they enter only where relevant to the recoil-order discussion.

Together, these tools provide a systematic basis for calculating recoil contributions to nuclear shape factors and for connecting those contributions to nuclear structure. Their improved treatment is important both for refining Standard-Model predictions and for distinguishing possible signatures of new interactions from ordinary recoil effects. As experimental precision advances, calculations with controlled nuclear-structure uncertainties will be essential to fully exploit $\beta$ decay as a probe of fundamental physics.

\begin{ack}[Acknowledgments]
T The authors thank Smarajit Triambak, Oscar Naviliat-Cuncic, Leendert Hayen, Kristina Launey and Nicholas Scielzo for useful discussions. This material is based upon work supported by the U.S. Department of Energy, Office of Science, Office of Nuclear Physics, under the FRIB Theory Alliance award DE-SC0013617, as well as work supported by the Israel Science Foundation Grant No. 1446/16 and a scholarship sponsored by the Ministry of Science and Technology, Israel.
\end{ack}

\section*{Appendix}
\subsection*{Derivation of allowed $\beta$ decay operators in the long-wavelength limit}

As we mentioned in Sec. \ref{sec:shape_allowed}, $\hat C_0^V$ and $\hat L_1^A$ at the leading order are associated with the Fermi and Gamow-Teller operators, respectively. Here we show that connection explicitly by applying the long-wavelength approximation ($qr\sim qR<<1$), which simplifies the spherical Bessel's functions to
\begin{equation}
j_J(qr) \xrightarrow[qr \to 0]{} \frac{(qr)^J}{(2J+1)!!}.
\end{equation}
Hence, for allowed Fermi transitions (no spin change, $J=0$), one obtains the familiar form of the operator
\begin{equation}
    \hat{C}_0^V\approx \frac{g_V}{\sqrt{4\pi}}\sum_{j=1}^A\tau_j^\pm, 
\end{equation}
where we used $Y_{00}\equiv1/\sqrt{4\pi}$. We note that all other vector multipole operators are typically much smaller than $\hat{C}_0^V$ in the long-wavelength limit as they are either suppressed by the $q/m_N$ factor or by powers of $qr$ from  the general asymptotic behavior of the spherical Bessel function $j_J(qr)$.

Similarly, for the axial operators in the long-wavelength limit, the lowest multipole order that mediates allowed Gamow-Teller transitions ($\Delta J = 1$, no parity change) is $J=1$. Applying the long-wavelength limit to the spherical Bessel function for $J=1$ yields
$j_1(qr) \approx \frac{qr}{3}$.
Substituting this into the definition of $S_{1M}(q\vec{r})$ gives:
\begin{equation}
S_{1M}(q\vec{r}) \approx \frac{qr}{3} Y_{1M}(\hat{r}).
\end{equation}
Using the standard relation connecting spherical harmonics to the spherical components of the position vector, $r Y_{1M}(\hat{r}) = \sqrt{\frac{3}{4\pi}} r_M$, we rewrite the function as:
\begin{equation}
S_{1M}(q\vec{r}) \approx \frac{q}{3} \sqrt{\frac{3}{4\pi}} r_M = \frac{q}{\sqrt{12\pi}} r_M.
\end{equation}
We now substitute this result into the vector divergence core of the operator. Because the spin vector $\vec{\sigma}$ acts strictly on the spin space and has no coordinate dependence ($\vec{\nabla} \cdot \vec{\sigma} = 0$), we apply the vector product identity $\vec{\nabla} \cdot (f \vec{A}) = (\vec{\nabla} f) \cdot \vec{A} + f(\vec{\nabla} \cdot \vec{A})$:
\begin{align}
\vec{\nabla} \cdot \left[ S_{1M}(q\vec{r}) \vec{\sigma} \right] &\approx  \frac{q}{\sqrt{12\pi}} \sigma_M,
\end{align}
where $\sigma_M$ is the $M$-th spherical component of the spin operator.
Substituting the evaluated divergence back into the full definition of the multipole operator yields
\begin{equation}
\hat{L}_{1M}^A(q) \approx \frac{i}{q} \sum_{j=1}^{A} \tau_j^\pm \left( \frac{q}{\sqrt{12\pi}} \sigma_{j,M} \right).
\end{equation}
The explicit momentum dependence $q$ in the numerator and denominator cancels exactly, reducing to the Gamow-Teller form:
\begin{equation}
\hat{L}_{1M}^A(q) \approx \frac{i}{\sqrt{12\pi}} \sum_{j=1}^{A} \tau_j^\pm \sigma_{j,M}.
\end{equation}

Following the same procedure, one can derive a leading-order Gamow-Teller form from the axial electric transverse operator $\hat E_J^A$:
\begin{equation}
    \hat{E}_{1M}^A(q) \approx \frac{i}{\sqrt{6\pi}} \sum_{j=1}^{A} \tau_j^\pm \sigma_{j,M}.
\end{equation}
All other terms or higher multipoles are either suppressed by the small $q/m_N$ factor or $qr$. Hence, if one neglects the recoil-order terms and plugs the approximate expressions of $\hat C_0^V$, $\hat L_1^A$ and $\hat{E}_{1}^A$ into the expression for the shape function Eq. \eqref{eq:multipole_exp}, then one obtains
\begin{align}
\Theta &\approx  \, (1 + \vec{\beta}\cdot\hat{\nu}) \left|\langle J_f || \hat{C}_0^V || J_i \rangle\right|^2  
 + \left(1 - \vec{\beta}\cdot\hat{\nu} + 2(\vec{\beta}\cdot\hat{q})(\hat{\nu}\cdot\hat{q})\right) \left|\langle J_f || \hat{L}_1^A || J_i \rangle\right|^2 
 + \left(1 - (\vec{\beta}\cdot\hat{q})(\hat{\nu}\cdot\hat{q})\right) \left|\langle J_f || \hat{E}_1^{A} || J_i \rangle\right|^2 \nonumber \\ 
 &=(1 + \vec{\beta}\cdot\hat{\nu}) \frac{1}{4\pi} g_V^2 B(F) + \left[ \left(1 - \vec{\beta}\cdot\hat{\nu} + 2(\vec{\beta}\cdot\hat{q})(\hat{\nu}\cdot\hat{q})\right) \frac{1}{12\pi} + \left(1 - (\vec{\beta}\cdot\hat{q})(\hat{\nu}\cdot\hat{q})\right) \frac{2}{12\pi} \right] g_A^2 B(GT) \nonumber \\
 & = \frac{1}{4\pi} \left[ g_V^2 B(F)(1 + \vec{\beta}\cdot\hat{\nu}) + g_A^2 B(GT)\left(1 - \frac{1}{3}\vec{\beta}\cdot\hat{\nu}\right) \right],
\end{align}
where $B(F) \equiv \left|\langle J_f || \sum_j \tau_j^\pm || J_i \rangle\right|^2$ and $B(GT) \equiv \left|\langle J_f || \sum_j \tau_j^\pm \vec{\sigma}_j || J_i \rangle\right|^2$. 

If one were to ignore the beta-neutrino angular correlations and assume isotropic $\beta$-decay, the orientation-independent rate could be obtained by integrating over the full solid angles of electron ($d\Omega_e$) and neutrino ($d\Omega_\nu$) emission:
\begin{equation}
\Gamma \propto \int \Theta(\vec{\beta}, \hat{\nu}) \, \frac{d\Omega_e}{4\pi} \frac{d\Omega_\nu}{4\pi}.
\end{equation}
Because the relative orientation vector averages out over complete spherical paths ($\int (\vec{\beta}\cdot\hat{\nu}) \, d\Omega_e \, d\Omega_\nu = 0$), the directional correlation terms vanish:
\begin{equation}
\int \Theta(\vec{\beta}, \hat{\nu}) \, \frac{d\Omega_k}{4\pi} \frac{d\Omega_\nu}{4\pi} = \frac{1}{4\pi} \left[ g_V^2 B(F) + g_A^2 B(GT) \right].
\end{equation}
Plugging this result into the rate expression in Eq. \eqref{eq:dec_rate} and ignoring the outer corrections ($C_\text{corr}=1$) yields the familiar textbook formula (see. \textit{e.g.}, \cite{wong2008introductory}) for the allowed $\beta$-decay transition rate:
\begin{equation}
d\Gamma = \frac{G_F^2 V_{ud}^2}{2\pi^3} F(Z, E_e) p_e E_e (E_0 - E_e)^2 \left[ g_V^2 B(F) + g_A^2 B(GT) \right] dE_e .
\end{equation}
This exercise shows how many more corrections need to be accounted for in the modern precision $\beta$ decay experiments that test fundamental symmetries.

\subsection*{Derivation of the weak magnetism in the long-wavelength limit}
In \cite{Holstein:1974zf}, in the elementary particle formulation for allowed $\beta$-decay, the weak magnetism form factor $b_\text{WM}$ captures the leading-order recoil correction ($q/m_N$) originating from the vector current. For an allowed Gamow-Teller transition ($\Delta J = 1, \Delta \pi = \text{no}$), this process is governed by the $J=1$ transverse magnetic vector multipole operator $\hat{M}_{1}^{V} (q)$, with expression given in Eq. \eqref{eq:multipoles1b}. 
The vector spherical harmonic $\vec Y_{111}(\hat{r})$ is defined via the orbital angular momentum operator $\vec l = -i(\vec r \times \vec \nabla)$ acting on the scalar spherical harmonic $Y_{1M}(\hat{r})$:
\begin{equation}
\vec Y_{111}(\hat{r}) = \frac{1}{\sqrt{1(1+1)}} \vec l Y_{1M}(\hat{r}) = \frac{1}{\sqrt{2}} \vec l Y_{1M}(\hat{r}).
\end{equation}
Using the relation between $Y_{1M}(\hat{r})$ and the spherical component of the position vector $\vec r_M = \sqrt{\frac{4\pi}{3}} r Y_{1M}(\hat{r})$, we write:
\begin{equation}
\vec Y_{111}^M(\hat{r}) = \frac{1}{\sqrt{2}} \left( \sqrt{\frac{3}{4\pi}} \frac{1}{r} \vec l \vec r_M \right) = \frac{1}{\sqrt{2}} \sqrt{\frac{3}{4\pi}} \frac{1}{r} \left[ -i (\vec r \times \vec \nabla) \vec r_M \right] = \frac{i}{\sqrt{2}}\sqrt{\frac{3}{4\pi}}\frac{1}{r} (\hat{e}_M \times \vec r).
\end{equation}
Where $\hat e_M$ is the unit spherical vector. Multiplying by the leading-order expansion of $j_1(qr) \approx qr/3$, the expression for $\vec S_{111}(q\vec r)$ for $qr \ll 1$ is:
\begin{equation}
\vec S_{111}^M(q\vec r) \approx \frac{qr}{3} \left[ \frac{i}{\sqrt{2}}\sqrt{\frac{3}{4\pi}}\frac{1}{r} (\hat{e}_M \times \vec r) \right] = i \frac{q}{2\sqrt{6\pi}} (\hat{e}_M \times \vec r).
\end{equation}
Substituting our exact expression for $\vec S_{111}^M(q\vec r)$ into the first term inside the operator braces:
\begin{equation}
g_V \vec S_{111}^M(q\vec r_j) \cdot \vec \nabla_j = g_V \left[ i \frac{q}{2\sqrt{6\pi}} (\hat{e}_M \times \vec r_j) \right] \cdot \vec \nabla_j.
\end{equation}
Utilizing the scalar triple product vector identity $(\vec A \times \vec B) \cdot \vec C = \vec A \cdot (\vec B \times \vec C)$:
\begin{equation}
(\hat{e}_M \times \vec r_j) \cdot \vec \nabla_j = \hat{e}_M \cdot (\vec r_j \times \vec \nabla_j).
\end{equation}
Since $\vec l_j = -i(\vec r_j \times \vec \nabla_j)$, it follows that $(\vec r_j \times \vec \nabla_j) = i \vec l_j$. Substituting this back yields:
\begin{equation}
g_V \vec S_{111}^M(q\vec r_j) \cdot \vec \nabla_j = i \frac{q g_V}{2\sqrt{6\pi}} \hat{e}_M \cdot (i \vec l_j) = -\frac{q g_V}{2\sqrt{6\pi}} (\vec l_j)_M .
\end{equation}
For the second term, we take the curl of the vector field using the exact differential operation relation for vector spherical harmonics:
\begin{equation}
\vec \nabla \times [j_1(qr) \vec Y_{111}^M(\hat{r})] = i q \sqrt{\frac{2}{3}} j_0(qr) \vec Y_{101}^M(\hat{r}) + i q \sqrt{\frac{1}{3}} j_2(qr) \vec Y_{121}^M(\hat{r}).
\end{equation}
As $qr \ll 1$, $j_0(qr) \approx 1$ and $j_2(qr) \approx \frac{(qr)^2}{15} \rightarrow 0$. The rank-1, orbital-0 vector harmonic simplifies via Clebsch-Gordan reduction to:
\begin{equation}
\vec Y_{101}(\hat{r}) = Y_{00}(\hat{r})\hat{e}_M = \frac{1}{\sqrt{4\pi}}\hat{e}_M.
\end{equation}
Thus, the curl behaves exactly as:
\begin{equation}
\vec \nabla \times \vec S_{111}(q\vec r) \approx i q \sqrt{\frac{2}{3}} (1) \left(\frac{1}{\sqrt{4\pi}}\hat{e}_M\right) = i q \frac{1}{\sqrt{6\pi}} \hat{e}_M.
\end{equation}
Plugging this expression back into the spin term of the multipole operator:
\begin{equation}
\frac{i}{2}\mu \left[ \vec\nabla_j \times \vec S_{111}(q\vec r_{j}) \right] \cdot \vec\sigma_{j} = \frac{i}{2}\mu \left( i q \frac{1}{\sqrt{6\pi}} \hat{e}_M \right) \cdot \vec \sigma_j = -\frac{q \mu}{2\sqrt{6\pi}} (\vec \sigma_j)_M.
\end{equation}
Combining both parts into $\hat{M}_{1}^{V}(q)$ yields:
\begin{align}
\hat{M}_{1M}^{V}(q) &= -\frac{i}{m_{N}} \sum_{j=1}^{A}\tau_{j}^{\pm} \left\{ -\frac{q g_V}{2\sqrt{6\pi}} (\vec l_j)_M - \frac{q \mu}{2\sqrt{6\pi}} (\vec \sigma_j)_M \right\} \nonumber \\
&=\frac{i q}{2\sqrt{6\pi} m_N} \sum_{j=1}^{A}\tau_{j}^{\pm} \left[ g_V \vec l_j + \mu \vec \sigma_j \right]_M.
\end{align}
To extract the form factor, we evaluate the nuclear reduced matrix element $\langle J_f \Vert \hat{M}_{1}^V(q) \Vert J_i \rangle$:
\begin{equation}
\langle J_f \Vert \hat{M}_{1}^V(q) \Vert J_i \rangle = \left(\frac{i q}{2\sqrt{6\pi} m_N}\right) \langle J_f \Vert \sum_{j=1}^A \tau_j^\pm (g_V \vec l_j + \mu \vec \sigma_j) \Vert J_i \rangle .
\end{equation}
Comparing with the form of $b_\text{WM}$ in Eq. \eqref{eq:Holstein} yields:
\begin{equation}
b_{\mathrm{WM}} \approx \frac{(-1)^{J_i-J_f}}{\sqrt{2J_i+1}} \left(\frac{2\sqrt{6\pi} A m_N}{i q} \right)\langle J_f \Vert \hat{M}_{1}^V(q) \Vert J_i \rangle .
\end{equation}


\bibliographystyle{Harvard}
\bibliography{encyclopedia}

\end{document}